\documentclass[letterpaper,twocolumn,preprintnumbers,superscriptaddress,nofootinbib,number]{revtex4}
\usepackage{graphicx}
\usepackage{color}
\usepackage{bm}
\usepackage{amsmath,amssymb,url,hyperref}
\newcommand\ignore[1]{}
\def\bbar{{\overline{b}b}}
\def\qqbar{{Q\overline{Q}}}

\newcommand{\bss}[1]{\ensuremath{{\boldsymbol{#1}}}}
\usepackage{placeins}

\def\sfrac#1#2{{\textstyle\frac{#1}{#2}}}

\begin{document}

\preprint{MIT-CTP 4417}
 \title{Quarkonium
  at non-zero isospin density} \author{William Detmold} \author{Stefan
  Meinel} \affiliation{Center for Theoretical Physics, Massachusetts
  Institute of Technology, Cambridge, MA 02139, USA} \author{Zhifeng
  Shi}\affiliation{Department of Physics, The College of William \&
  Mary, Williamsburg, VA 23187, USA} \affiliation{Jefferson Lab,
  Newport News, VA 23606, USA }

\begin{abstract}
  We calculate the energies of  quarkonium bound states
  in the presence of a medium of nonzero isospin density using lattice
  QCD. The medium, created using a canonical (fixed isospin charge)
  approach, induces a reduction of the quarkonium
  energies. As the isospin
  density increases, the energy shifts first increase and then
  saturate. The saturation occurs at an isospin density close to that where
  previously a qualitative change in the behaviour of the energy density of the medium has been 
  observed, which was conjectured to correspond to a transition from a pion gas to a Bose-Einstein
  condensed phase. The reduction of the quarkonium energies becomes more
  pronounced as the heavy-quark mass is decreased, similar to the
  behaviour seen in two-colour QCD at non-zero quark chemical
  potential. In the process of our analysis, the $\eta_b$-$\pi$ and $\Upsilon$-$\pi$
  scattering phase shifts are determined at low momentum. An interpolation of the scattering lengths to the physical pion
  mass gives $a_{\eta_b,\pi} = 0.0025(8)(6)$ fm and $a_{\Upsilon,\pi} = 0.0030(9)(7)$ fm.
\end{abstract}
\maketitle

\section{Introduction}

An important probe of exotic phases of QCD matter is the way in which
 heavy quarkonium  propagation is modified by the presence of that
 matter.
  The heavy
quarks can in some sense be viewed as separable from the medium which
is predominantly composed of light quark and gluonic degrees of
freedom.  At non-zero temperature, the suppression of the propagation
of $J/\psi$ particles is a key signature for the formation of a
quark-gluon plasma \cite{Matsui:1986dk}. This suppression has been
observed for charmonium in various experiments at SPS, RHIC and the
LHC and recently in the $\Upsilon$ spectrum at the LHC
\cite{Chatrchyan:2011pe}.  Quarkonium propagation is naturally also
expected to be a sensitive probe of other changes of phase such as
those that occur at high density or large isospin density.

Since the effects of QCD matter on quarkonia are essentially
non-perturbative in origin, a systematic evaluation requires input
from lattice QCD. At some level, these effects can be distilled to a
change in the potential between the quark--anti-quark ($\qqbar$) pair
that binds them into quarkonium. At non-zero temperature but zero
density, this has been studied extensively using lattice QCD (see
Ref.~\cite{Petreczky:2012rq} for a recent overview) where strong
screening effects are seen near the deconfinement scale. Significant effects
are also seen in investigations of the properties of charmonium
and bottomonium spectral functions at non-zero temperature (see
\cite{Petreczky:2012rq,Aarts:2012ka}).

Modifications of the potential or quarkonium properties 
will also occur for non-zero density.
Ref.~\cite{Detmold:2008bw} has
investigated the static potential in the presence of a gas of pions
and below we briefly address how the medium affects the binding of
quarkonium through solving the Schr\"odingier equation for the modified
potential.  As the main focus of this work, however, we explore the
effects of isospin charge density on quarkonium bound state energies
more directly by using lattice NRQCD (non-relativistic QCD) to compute
quarkonium correlation functions in the presence of a medium of
varying isospin chemical potential. At low isospin densities, and
correspondingly low chemical potentials, we find that the ground state
energy of the quarkonium systems decreases with increasing density,
showing qualitative agreement between the potential model calculation and the
QCD calculation. However, at an effective isospin chemical potential
$\mu_I \sim \mu_{I,{\rm peak}} = 1.3\ m_\pi$ (where previous
calculations of the energy density of the isospin medium have
suggested a transition to a Bose-Einstein condensed state
\cite{Detmold:2012wc} in line with theoretical expectations \cite{Son:2000xc}), the
effect of the medium on the quarkonium energy appears to saturate to a
constant shift. At large isospin densities, the determination of the
energy shift becomes statistically noisy.

Our study is presented as follows. In Section \ref{sec:pot}, we
construct an isospin density dependent potential model of quarkonium
states, and calculate the expected shift of the quarkonium energies in
the medium. Section \ref{sec:latt} presents the methodology of the
lattice QCD calculation of these energy shifts, with results presented
in Sections \ref{sec:UpsilonPi} and \ref{sec:res}. We conclude with a
discussion of the results and possible extensions to the current work.

\section{Modification of the static quark--anti-quark potential}
\label{sec:pot}

\begin{figure*}[!tbp]
  \begin{center}
    \includegraphics[width=\textwidth]{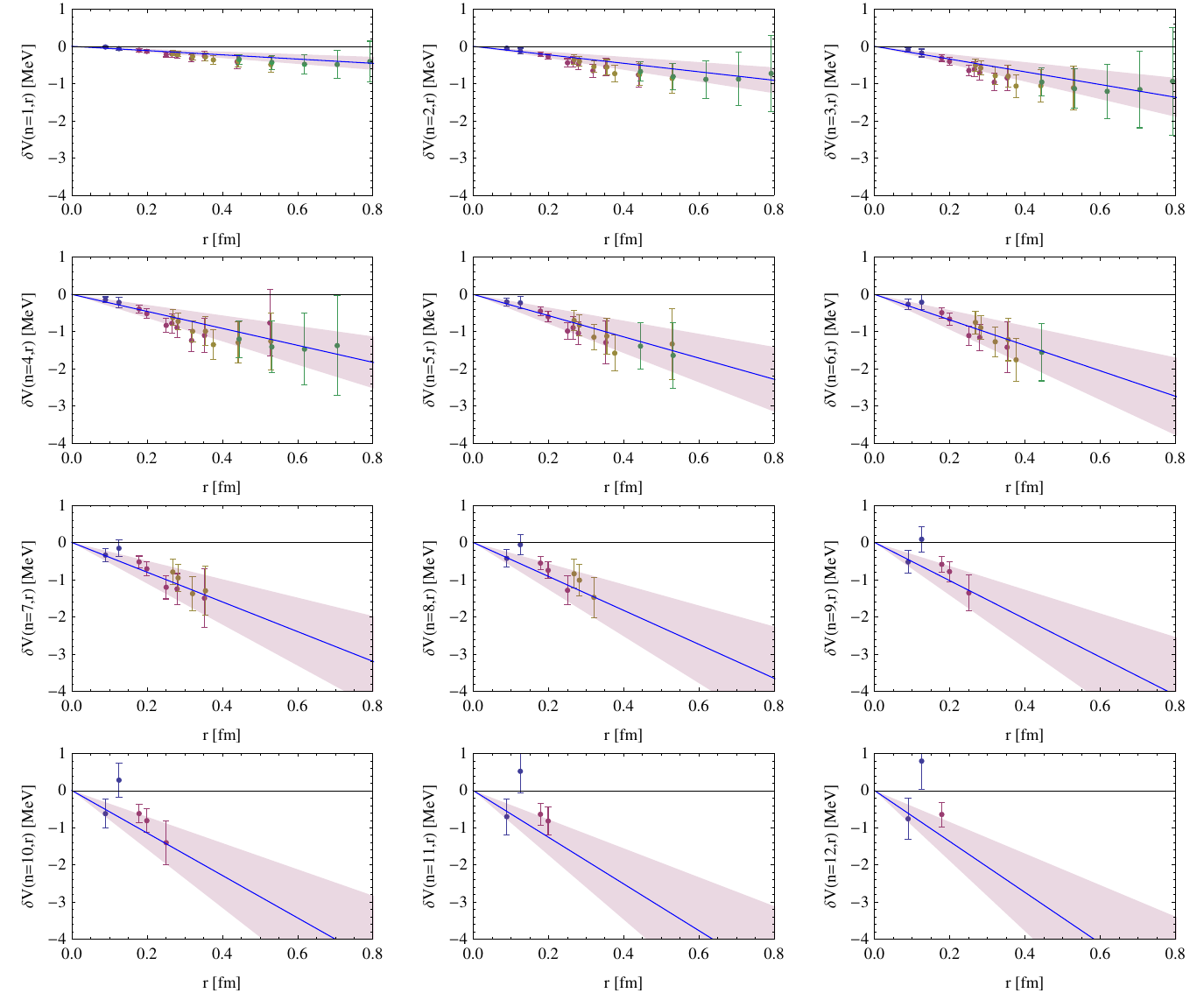}
    \caption{Shifts in the static potential computed in
      Ref.~\protect\cite{Detmold:2008bw} fitted to the simple 
    form discussed in the text.}
    \label{fig:staticpot}
  \end{center}
\end{figure*}

Here we briefly review the results obtained in
Ref.~\cite{Detmold:2008bw} on the screening of the static 
potential by a pion gas and ascertain the expected effects of these
modifications on quarkonium spectroscopy.  Ref.~\cite{Detmold:2008bw}
presented a sophisticated calculation of the static quark--anti-quark
potential in a pionic medium as a function of the separation of the
$Q\overline{Q}$ pair, $r$, and as a function of the isospin charge,
$n$, of the medium. This required measurement of Wilson loops of
various extents in space and in time and to obtain signals for large
loops, different levels of  HYP (hypercubic) smearing  \cite{Hasenfratz:2001hp} were used
in overlapping regions of $r$, complicating the analysis.  As will be
discussed below, constructing appropriate ratios of correlation
functions was also critical in order to obtain statistically clean
measurements. The central results of this work were that the potential
is screened by the presence of the medium and that this screening
effect is small. For the relatively low pion densities investigated in Ref.~\cite{Detmold:2008bw}, 
the dominant effect corresponded to a change in the
potential in the linearly rising region that was approximately
linearly dependent on both $r$ and $n$.  This form, $\delta
V(\rho_I,r)=\alpha\ \rho_I \ r$, corresponds to the physical
expectation of a gas of weakly-interacting pions permeating a
flux-tube of constant radius between the static quark and 
anti-quark, and is a picture in which the appearance of the isospin
density, $\rho_I $, is natural.  Performing a correlated fit to the results presented in
Ref.~\cite{Detmold:2008bw} using this form, we are able to describe
the data well, as shown in Fig.~\ref{fig:staticpot}, and find that
$\alpha=-8(3)$ MeV fm$^2$. This result is for a pion mass of $m_\pi \sim 320$ MeV \cite{Detmold:2008bw}.

\begin{figure}[!tbp]
  \begin{center}
    \includegraphics[width=0.9\columnwidth]{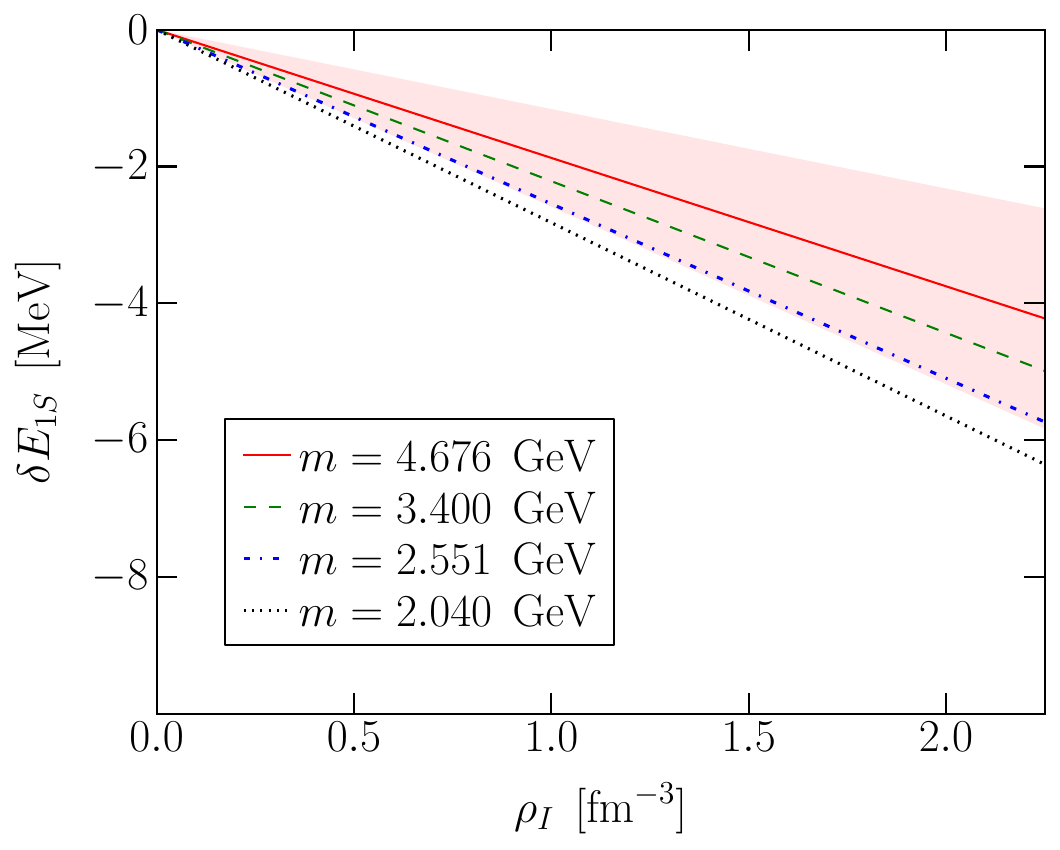}\\
    \includegraphics[width=0.9\columnwidth]{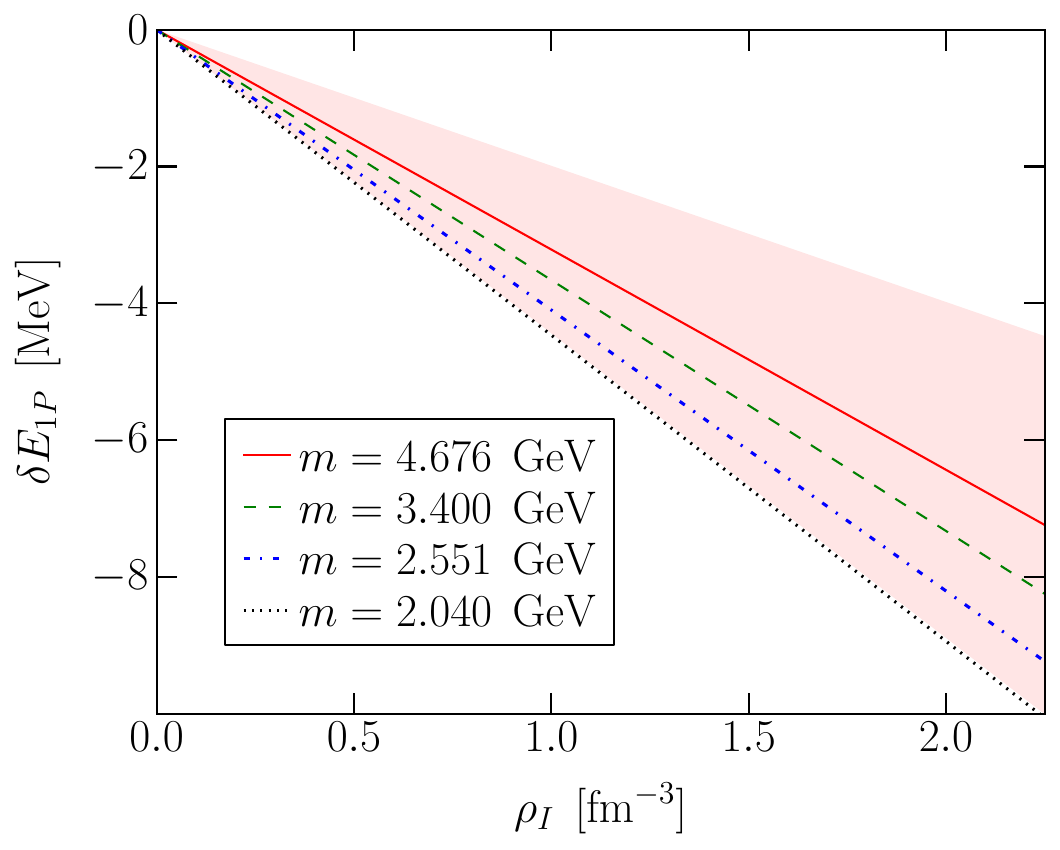}
    \caption{Shifts in the energies of the $1S$ (upper) and $1P$
      (lower) states in quarkonium as a function of the isospin
      density, computed in a potential model. Results are shown for four different
      values of the heavy-quark mass with the uncertainty shown only
      for the mass closest to the physical bottom-quark mass, $m=4.676$~GeV (uncertainties for
      the other masses are of similar size).}
    \label{fig:potmodel}
  \end{center}
\end{figure}

To estimate the effects on quarkonium spectroscopy, we use the
Cornell potential $V_{\rm Cornell}(r)=-(4/3)\alpha_s / r + \kappa\:r$
with $\alpha_s = 0.24$ and $\sqrt{\kappa}=468$ MeV (values fixed in
vacuum from Ref.~\cite{Bali:1997am}) and augment it with the small
screening shift discussed above.  We then solve the radially symmetric
Schr\"odinger equation numerically for angular momentum $\ell$ and reduced mass
$m_{\rm red}=m/2$ (where $m$ is the heavy-quark mass),
\begin{eqnarray}
  \label{eq:schrodigner}
  \Big[-\frac{1}{2m_{\rm red}}\frac{d^2}{dr^2}+\frac{\ell(\ell+1)}{2m_{\rm red}\ r^2} 
  \hspace*{4cm} 
  \\ \nonumber
  \hspace*{1cm} 
  +V_{\rm Cornell}(r) +\delta V(\rho_I,r)\Big]u_\ell(r)=E\ u_\ell(r)\,,
\end{eqnarray}
to establish the wave functions and eigenstate energies for the
various quantum numbers.  The energy shift is then defined
simply as the difference of the resulting energy from that where
$\delta V(\rho_I,r)$ is omitted. We calculate this shift for both the
$1S$ and $1P$ states and various different values of the heavy-quark
mass as shown in Fig.~\ref{fig:potmodel}.  We note that one could use
only the Cornell potential to determine the wave functions and include
the additional small shift from the screening as a perturbation,
calculating
\begin{eqnarray}
  \delta E(\rho_I) \sim \int d r \ u^{(0)\ast}_{\ell}(r) \delta V(\rho_I,r) u^{(0)}_{\ell}(r)\,,
\end{eqnarray}
where $u_\ell^{(0)}(r)$ are the solutions to Eq.~(\ref{eq:schrodigner})
when $\delta V(\rho_I,r)$ is omitted.  As the shift is small, we
expect this will give consistent results.

 Since $P$-wave states 
are more extended in size, they probe regions of the potential where the 
shift is larger and consequently  we find that the energy shift is larger for these 
states than for the $S$-wave states. The effect also increases as the heavy-quark mass
decreases, again because of the larger size of the lighter systems.

In the following, we determine quarkonium eigen-energies in (NR)QCD at
non-zero isospin density and investigate to what extent
they are predicted by the potential model described above based on a
screening pion gas.

\section{Lattice methodology}
\label{sec:latt}

\subsection{Lattice details}

In this study, we make use of anisotropic gauge configurations
generated by the Hadron Spectrum and NPLQCD collaborations. The full
details of the action and algorithms used to generate the
configurations are discussed in the original works, Refs.~\cite{Edwards:2008ja,Lin:2008pr};
here we summarise the salient features of the configurations and the
measurements that we perform.  A tree-level, tadpole-improved gauge action
\cite{Luscher:1984xn}, and $n_f=2+1$ flavour clover fermion
action~\cite{Sheikholeslami:1985ij} are used. Two levels of stout
smearing~\cite{Morningstar:2003gk} with weight $\rho = 0.14$ are applied in spatial
directions only in order to preserve the ultra-locality of the action
in the temporal direction. The gauge action is constructed without a
$1\times2$ rectangle in the time direction for the same reason. In this study, we
make use of a single spatial lattice spacing, $a_s=0.1227(8)\ {\rm
  fm}$~\cite{Lin:2008pr} and have a renormalised anisotropy of
$\xi=a_s/a_t=3.5$, where $a_t$ is the temporal lattice spacing. We also
work at a single value of the light-quark mass for this
exploratory investigation and use a strange-quark mass that is close to its physical
value; these values correspond to a pion mass of $m_\pi \sim
390$~MeV and a kaon mass of $m_K\sim 540$ MeV. For these parameters,
we investigate three different ensembles, corresponding to different
physical volumes and temporal extents as shown in Table
\ref{tab:lattices}.  The different physical volumes allow us to access
a large range of isospin densities in our study, and the different
temporal extents provide control of thermal effects as discussed in
Ref.~\cite{Detmold:2012wc}.  On these gauge configurations
we calculate correlation functions involving light quarks and use the
colourwave propagator basis introduced in Ref.~\cite{Detmold:2012wc}, fixing to Coulomb
gauge and using plane-wave sources and sinks for a range of low momenta
($N_{\rm mom}$ in total on each ensemble, see
Table~\ref{tab:lattices}). For each case, we calculate light-quark propagators on
$N_{\rm cfg}$ configurations from $N_{\rm src}$ time-slices, equally
spaced throughout the temporal extent. Details of the NRQCD heavy quark 
propagator calculations are discussed below.

\begin{table}[h!]
  \begin{tabular}{ccccccccccccccccc}
    \hline\hline
    $N_s^3\times N_t$ &  \hspace{-0.2ex} &  $L[{\rm fm}]$ &  \hspace{-0.2ex} & $m_{\pi}
    L$ &  \hspace{-0.2ex} & $m_{\pi} T$ &  \hspace{-0.2ex}  &  $u_{0s}$  & \hspace{-0.2ex} &  $N_{\rm cfg}$& \hspace{-0.2ex} &  
    $N_{\rm src}$  & \hspace{-0.2ex} & $N_{\rm mom}$ \\
    \hline
    $16^3\times 128$ && 2.0 && $3.86$ && $8.82$  && $0.7618$ && 334 && 8 && 33\\
    $20^3\times 256$ && 2.5 && $4.82$ && $17.64$  && $0.7617$ && 170 && 16 && 7\\
    $24^3\times 128$ && 3.0 &&  $5.79$ && $8.82$ && $0.7617$ && 170 && 8 && 19 \\
    \hline\hline
  \end{tabular}
  \caption{\label{tab:lattices}Details of the ensembles and measurements used in this work.
  $u_{0s}$ is defined as the fourth root of the spatial plaquette.}
\end{table}

\subsection{Multi-pion lattice correlators}
  
In order to produce the medium that will modify the propagation of the
quarkonium states, we use the canonical approach of constructing 
many-pion correlation functions that is described in detail in
Ref.~\cite{Detmold:2012wc}, using methods developed there and in
earlier works
\cite{Beane:2007es,Detmold:2008fn,Detmold:2008yn,Detmold:2010au,Detmold:2011kw}.  As
discussed therein, correlators of a fixed isospin charge, ${
  n}=\sum_{i=1}^N n_i $, and total momentum, ${\bf P}_f$, making use
of $N$ sources, are given by \def\bx{{\bf x}} \def\bxp{{\bf x}^\prime}
\def\by{{\bf y}} \def\byp{{\bf y}^\prime} \def\bp{{\bf p}}
\def\bpp{{\bf p}^\prime}
\begin{eqnarray}
  C_{n_1,\cdots, n_N}(t,{\bf P}_f)
  &=&  \langle {\cal O}_{n\pi^+}(t){\cal O}_{n\pi^+}^\dagger(0)\rangle
  \nonumber
  \\
  &\hspace*{-4cm}=& 
  \hspace*{-2cm}\left \langle \prod_{i=1}^N\left( \sum_{{\bx}_i,{\bx}^\prime_i} 
      e^{-i({\bp}^i_1 {\bx}_i-{\bp}^i_2 {\bx}^\prime_i)}
      {\overline d}({\bx}^\prime_i,t)\gamma_5 u({\bx}_i,t)\right)^{n_i} \right. 
   \nonumber \\
   &&\hspace*{-1.5cm}\times
  \left. \prod_{j=1}^{n}\left(\sum_{{\by}_j} e^{i{\bp}_{f_j} {\by}_j}{\overline u}
      ({\by}_j,0)\gamma_5 d({\by}_j,0)\right) \right \rangle,
  \label{eq:C_m_sources_mom}
\end{eqnarray}
where ${\bf P}_f=\sum_{i=1}^{ n}{\bp}_{f_i}$,  and, for momentum conservation,
$\sum_{i=1}^N({\bp}^i_1 - {\bp}^i_2) = \sum_{j=1}^{
  n}{\bp}_{f_j}$. In what follows, we will set ${\bf P}_f=0$ but the
${\bp}_{f_i}$ and $\bp^{i}_{1,2}$ take various values subject to these
constraints; different choices of the momenta defining the
interpolating operators will have different overlap onto the
eigenstates of the chosen ${\bf P}_f$ but provide additional
statistical resolution in the determination of the energy of the
system.

\begin{figure}[!t]
  \centering
  \includegraphics[width=0.95\columnwidth]{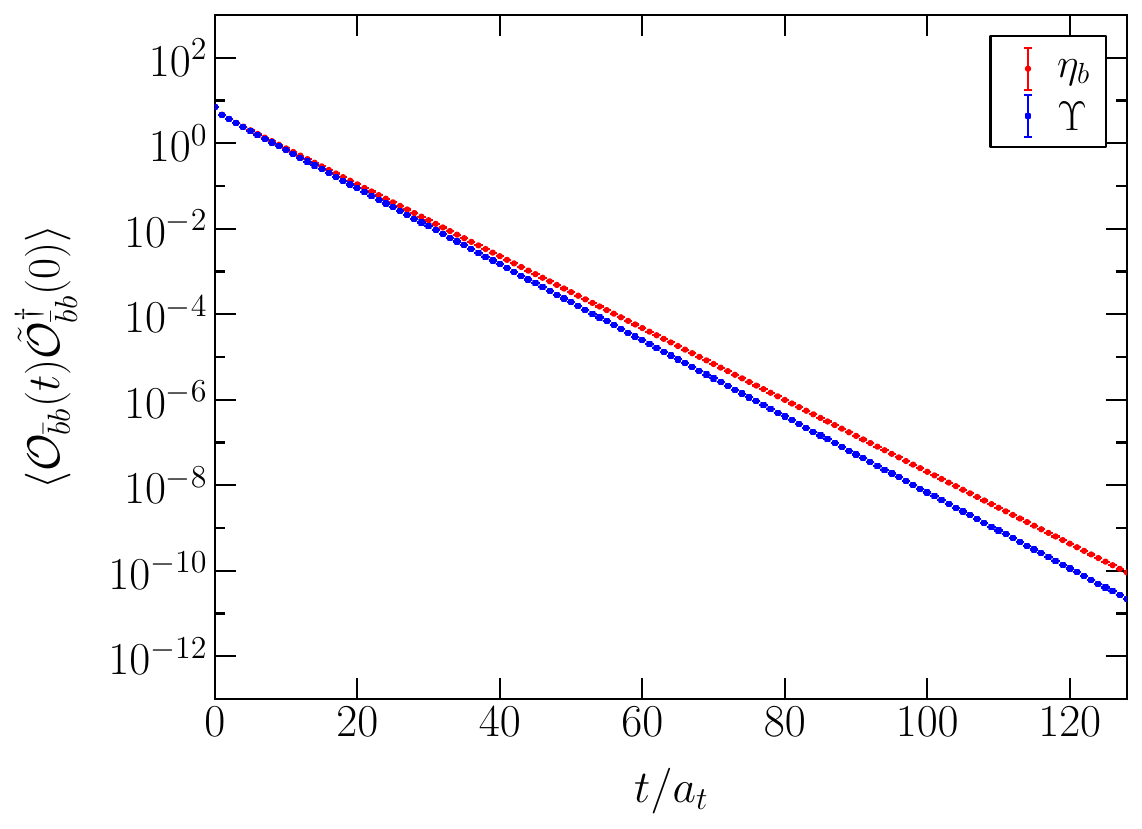}\\ \vspace*{3mm}
  \includegraphics[width=0.95\columnwidth]{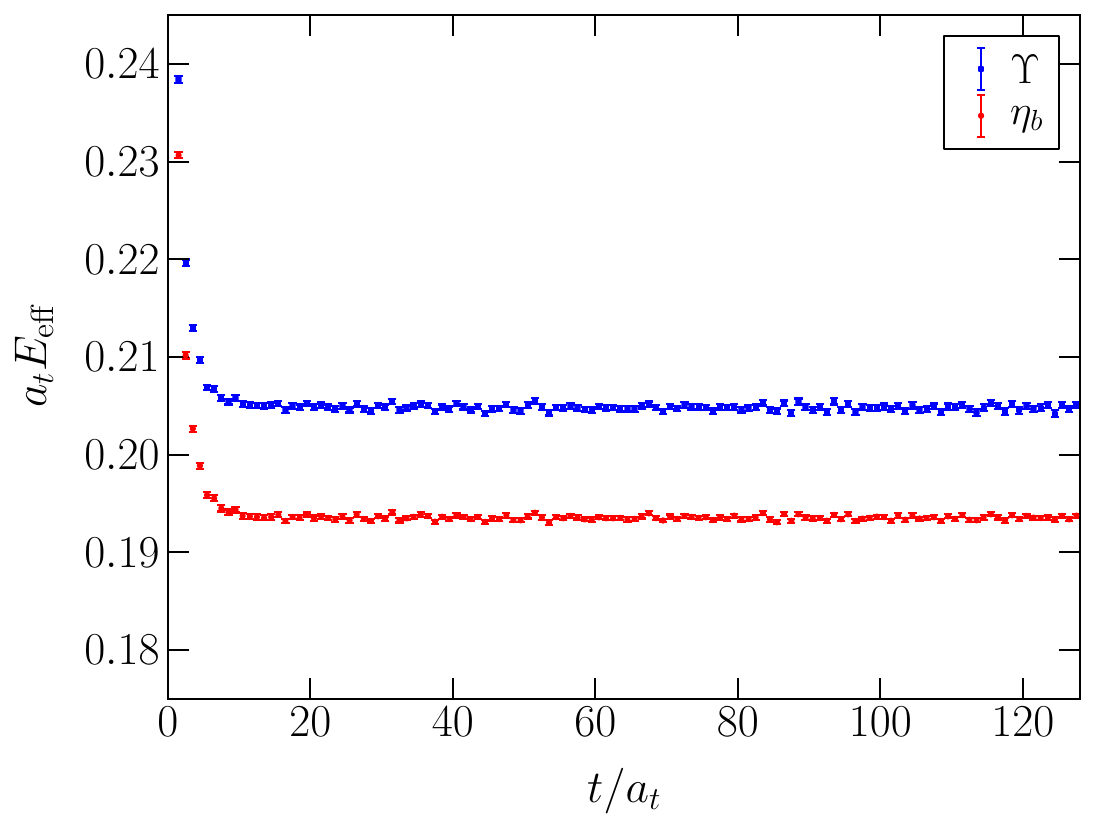}
  \caption{\label{fig:etabUpsiloneffmass}$\eta_b$ and $\Upsilon$
    correlators (upper) and effective energies (lower) on the
    $20^3\times 256$ ensemble, for $a_s m=2.75$.}
\end{figure}
\begin{figure}[!h]
  \centering
  \includegraphics[width=0.95\columnwidth]{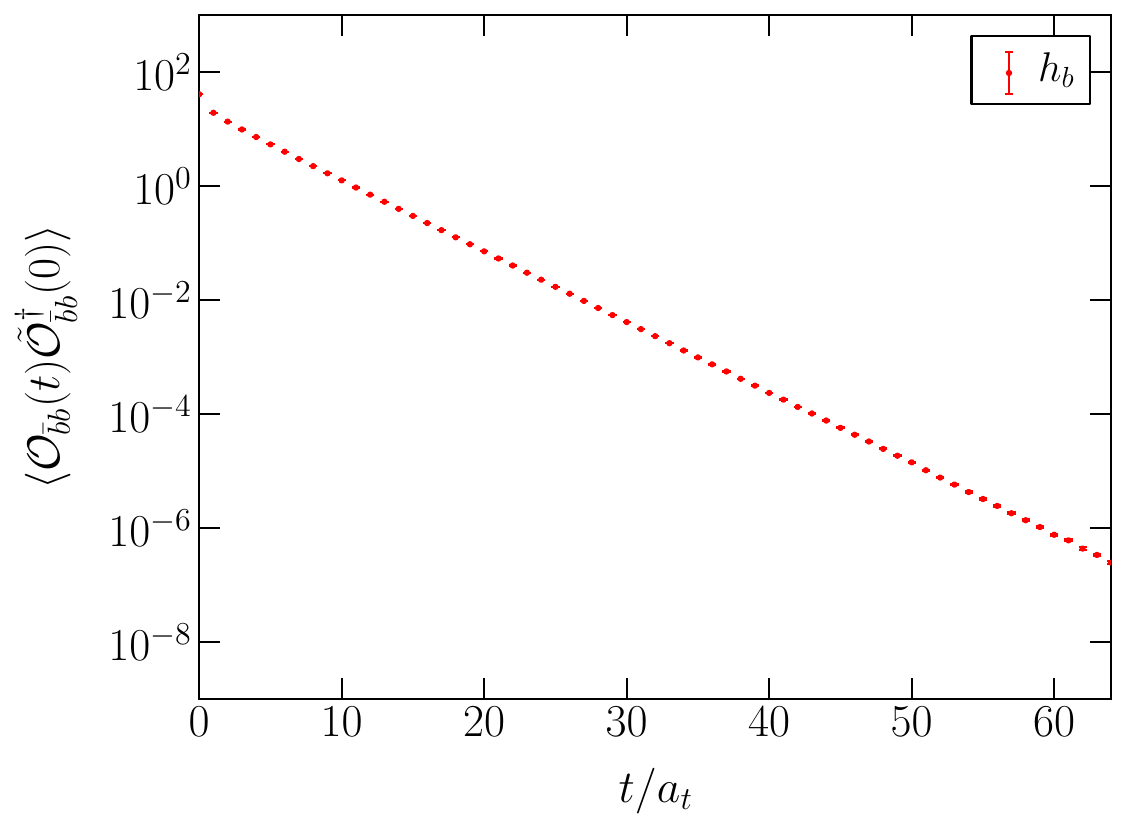}\\ \vspace*{3mm}
  \includegraphics[width=0.95\columnwidth]{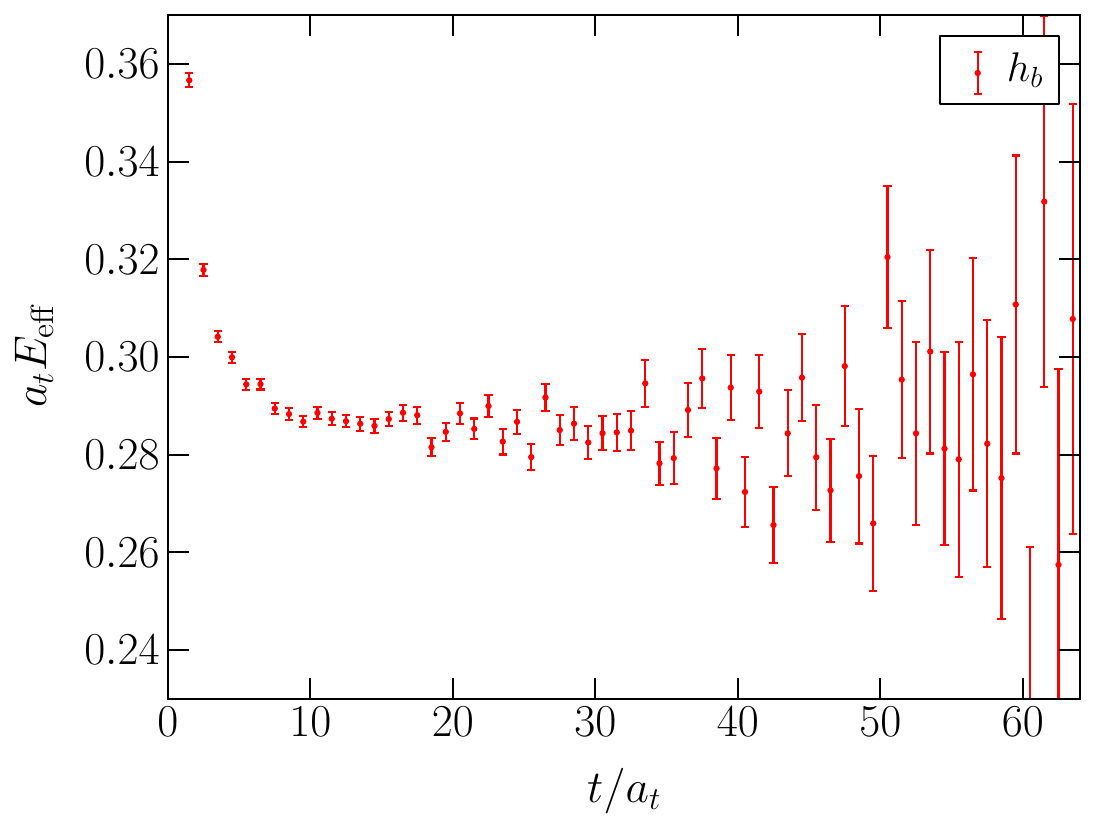}
  \caption{\label{fig:heffmass}$h_b$ correlator (upper) and effective
    energy (lower) on the $16^3\times 128$ ensemble, for $a_s m=2.75$.}
\end{figure}

To construct these correlation functions, we work in Coulomb gauge and compute light-quark
colourwave propagators
\begin{eqnarray}
  S_{u/d}({\bp},t;{\bpp},0) = \sum_{\bf x}e^{-i{\bf p}{\bf x}}S_{u/d}({\bf x}, t; {\bf p}^{\prime},0),
\end{eqnarray}
where
\begin{eqnarray}
  S_{u/d}({\bf x},t; {\bf p}^{\prime},0) = \sum_{\bf y}e^{i{\bf p}^{\prime}{\bf y}}S_{u/d}({\bf x}, t; {\bf y}, 0)  \nonumber
\end{eqnarray}
is a solution of the lattice Dirac equation:
\begin{eqnarray}
  \sum_{{\bf x},t}D({\bf y}, \tilde{t}; {\bf x},t)S_{u/d}({\bf x},t; {\bf p}^{\prime},0) = e^{i{\bf p}^{\prime}{\bf y}}\delta_{\tilde{t},0}\,. \nonumber
\end{eqnarray}
The contractions implicit in Eq.~(\ref{eq:C_m_sources_mom}) can be
written in terms of a matrix $\tilde A$, the $12\times12$ sub-blocks
of which are given by
\begin{eqnarray}
  {\tilde A_{k,i}}\left(t\right) &=& \sum_{\bp}
  S\left({\bp}^k_1,{\bp}\right)S^\dagger\left(-{\bp}^i_2,{\bp}_{f_i}-{\bp}\right), 
\end{eqnarray}
where $k,i$ label the source and sink, and the dependence on
${\bp}_1^k$, ${\bp}_2^i$, and ${\bf P}_f$ is suppressed.  The correlators
above can be extracted by noting that combinations of the
$C_{n_1,\cdots, n_N}$ for a given ${n}=\sum_{i=1}^N n_i $ are
the coefficients of the expansion of
\begin{eqnarray}
  \det[1+\lambda \tilde A] = 1 + \lambda C_{1\pi} + \lambda^2 C_{2\pi} + \cdots +
  \lambda^{12N} C_{12N\pi}\,,
  \label{equ:det_expansion} 
\end{eqnarray}
 and can be
computed efficiently using the methods of Ref.~\cite{Detmold:2012wc}.
The different $C_{n_1,\cdots, n_N}$ for a given $n$ occur in complicated 
combinations in this expansion, however we are explicitly only  interested in the energies of the
system, so the particulars of the combination are irrelevant.

These correlators have been studied in detail in  previous work~\cite{Detmold:2012wc} and
we do not present them again here.  As investigated in detail in
Ref.~\cite{Detmold:2012wc}, many-pion correlations contain thermal
contributions in which parts of the system propagate around the
temporal boundary.  In our choice of fitting ranges in the analysis presented
below, we are careful to remain away
from the regions in Euclidean time that are contaminated by either
excited states or by these thermal effects.

\subsection{NRQCD for quarkonium correlators }
\label{sec:nrqcd}

To implement the heavy quarks in our quarkonia systems, we use a
lattice discretisation of non-relativistic QCD (NRQCD).  Since our
light quark and gluon degrees of freedom are defined on an
anisotropic lattice, we require lattice NRQCD \cite{Thacker:1990bm,
  Lepage:1992tx} formulated on an anisotropic lattice as first set out
in Ref.~\cite{Drummond:1998fd}. As the non-relativistic nature of
the theory already separates space and time, using a temporal lattice
spacing that differs from the spatial lattice spacing is a
very natural choice for NRQCD. Anisotropic lattice NRQCD has been used
for example to calculate the spectrum of quarkonium hybrid states
\cite{Manke:1998qc, Drummond:1999db}, and recently also to study
quarkonium at non-zero temperature \cite{Aarts:2011sm,Aarts:2012ka}.

The Euclidean action for the heavy quark field, $\psi$,
can be written as
\begin{equation}
  S_{\psi}=a_s^3\sum_{\mathbf{x},t}\psi^\dagger(\mathbf{x},t)\big[{\psi}(\mathbf{x},t)
  -K(t) \: {\psi}(\mathbf{x},t-a_t) \big], \label{eq:latact}
\end{equation}
where $K(t)$ is the operator that evolves the heavy-quark Green
function forward one step in time. Here we use the form
\begin{eqnarray}
  K(t)&=&\left(1-\frac{a_t\:\delta H|_t}{2}\right)
  \left(1-\frac{a_t H_0|_t}{2n} \right)^n U_0^\dag(t-a_t)
  \hspace*{1cm}
  \nonumber\\
  &&\hspace*{0.5cm}
 \times\left(1-\frac{a_t H_0|_{t-a_t}}{2n} \right)^n
  \left(1-\frac{a_t\:\delta H|_{t-a_t}}{2}\right)\,,
  \label{eq:mNRQCD_action_kernel}
\end{eqnarray}
where $U_0$ are the temporal gauge links.  In this expression, 
\begin{equation}
  H_0 = -\frac{\Delta^{(2)}}{2 m}, \label{eq:H0}
\end{equation}
is the order-$v^2$ term in the NRQCD velocity expansion, and $\delta H$ is
a correction term given by
\begin{eqnarray}
  \nonumber
  \delta H&=&-c_1\:\frac{\left(\Delta^{(2)}\right)^2}{8 m^3}+c_2\:\frac{ig}{8 m^2}\:\Big(\bss{\nabla}\cdot\mathbf{\widetilde{E}}
  -\mathbf{\widetilde{E}}\cdot\bss{\nabla}\Big)\\
  \nonumber&&-c_3\:\frac{g}{8 m^2}\:\bss{\sigma}\cdot
  \left(\bss{\widetilde{\nabla}}\times\mathbf{\widetilde{E}}
    -\mathbf{\widetilde{E}}\times\bss{\widetilde{\nabla}} \right)-c_4\:\frac{g}{2 m}\:\bss{\sigma}\cdot\mathbf{\widetilde{B}}\\
\nonumber  && + c_5\:\frac{a_s^2\Delta^{(4)}}{24m}
  -c_6\:\frac{a_t\left(\Delta^{(2)}\right)^2}{16n\:m^2} \\
\nonumber &&-c_7\:\frac{g}{8 m^3}\Big\{ \Delta^{(2)}, \: \bss{\sigma}\cdot\mathbf{\widetilde{B}} \Big \} \\
\nonumber &&
-c_8\:\frac{3g}{64 m^4}\left\{ \Delta^{(2)}, \: \bss{\sigma}\cdot \left(\bss{\widetilde{\nabla}}
\times\mathbf{\widetilde{E}}-\mathbf{\widetilde{E}}\times\bss{\widetilde{\nabla}} \right) \right\} \\
 &&
-c_9\:\frac{i g^2}{8 m^3}\:\bss{\sigma}\cdot(\mathbf{\widetilde{E}}\times\mathbf{\widetilde{E}})
  \label{eq:dH_full}
\end{eqnarray}
(the notation is as in Ref.~\cite{Meinel:2010pv}).
The operators with coefficients $c_1$ through $c_4$ are the
relativistic corrections of order $v^4$, and the operators
with coefficients $c_7$ through $c_9$ are the spin-dependent
relativistic corrections of order $v^6$. The operator with coefficient
$c_5$ removes the order-$a_s^2$ discretization error of $H_0$, and the
operator with $c_6$ removes the leading order-$a_t$ error in the time
evolution. Four-fermion operators, which arise beyond tree-level in the matching to QCD, are not included.
We set the coefficients of the spin-dependent order-$v^4$
terms to $c_3=1.28$ and $c_4=1.05$ to achieve the best
possible agreement of the bottomonium $1P$ and $1S$
spin splittings in vacuum with the experimental values.
We use the tree-level values $c_i=1$ for the other matching coefficients.
For tadpole improvement \cite{Lepage:1992xa} of the derivatives and field strengths, we set
$u_{0s}$ equal to the 4th root of the spatial plaquette (see Table
\ref{tab:lattices}), and set $u_{0t}=1$.

To avoid instabilities in the time evolution with the operator in 
Eq.~(\ref{eq:mNRQCD_action_kernel}), the parameter $n$ must be chosen such
that $\mathrm{max}[a_t H_0/(2n)] < 2$ \cite{Lepage:1992tx}.  On an
anisotropic lattice, this requires
\begin{equation}
  n > 3a_t/(2a_s^2 m) = 3/(2 \xi a_s m) \label{eq:stability}
\end{equation}
(interactions with gluons weaken this requirement slightly
\cite{Lepage:1992tx}). In this work, we set the bare heavy-quark mass
to $a_s m=2.75$ (which is near the $b$ quark mass) as well as to the lower
values $a_s m=2.0, 1.5, 1.2$. Because we have $\xi=3.5$, a stability
parameter of $n=1$ is sufficient in all cases.

When using NRQCD, all quarkonium energies are shifted by an unknown
constant (which is approximately equal to two times the heavy-quark
mass). This shift is state-independent and cancels in energy
splittings as well as in differences between energies extracted at zero
and non-zero isospin density. For the purpose of tuning
the heavy-quark mass, we measure the kinetic masses of the $\eta_b$
and $\Upsilon$ states, defined as
\begin{equation}
  a_t M_{\rm kin} = \frac{(a_s\mathbf{p})^2/\xi^2-\left[a_t E(\mathbf{p})-
  a_t E(0)\right]^2}{2\left[a_t E(\mathbf{p})-a_t E(0)\right]}\,, \label{eq:mkin}
\end{equation}
with one unit of lattice momentum, $|\mathbf{p}|=2\pi/L$. The
spin-averaged values of the $1S$ kinetic masses, 
$\overline{M_{\rm kin}}=(3M_{\rm kin}^{\Upsilon}+M_{\rm kin}^{\eta_b})/4$ 
computed on the
$16^3\times128$ ensemble (at $\rho_I=0$), are given in
Table~\ref{tab:mkin}. As a check of discretization errors, we have
also calculated the kinetic masses using larger lattice momenta.
For example, the kinetic masses computed using $|\mathbf{p}|=2\cdot2\pi/L$
differ from those computed using $|\mathbf{p}|=2\pi/L$ by only 0.4\% at $a_s m=2.75$
and by 2\% at $a_s m=1.2$.

\begin{table}[h!]
  \begin{tabular}{llcll}
    \hline\hline
    $a_s m$ & \hspace{2ex} &  $a_t \overline{M_{\rm kin}}$  & \hspace{2ex} & $\overline{M_{\rm kin}}$ (GeV) \\
    \hline
    1.2   &&  0.7698(81)   && 4.333(54)  \\
    1.5   &&  0.9377(16)   && 5.277(36)  \\
    2.0   &&  1.2259(12)   && 6.900(46)  \\
    2.75  &&  1.6667(12)   && 9.380(62)  \\
    \hline\hline
  \end{tabular}
  \caption{\label{tab:mkin}Spin-averaged quarkonium kinetic masses on the $16^3\times128$ ensemble.}
\end{table}

In the main calculations of this work, we use zero-momentum smeared
quarkonium interpolating fields of the form
\begin{equation}
  \mathcal{O}_{\bar bb}(t) = \sum_\mathbf{y'} \sum_\mathbf{y} \chi^\dag(\mathbf{y'},t) \Gamma(\mathbf{y}-\mathbf{y'}) \psi(\mathbf{y},t) \label{eq:bbarsink}
\end{equation} 
at the sink and
\begin{equation}
  \tilde{\mathcal{O}}_{\bar bb}(0) = \sum_{\mathbf{x}} \chi^\dag(\mathbf{0},0) \Gamma(\mathbf{x}) \psi(\mathbf{x},0)
\end{equation}
at the source. Here, $\chi$ is the heavy anti-quark field and
$\Gamma(\mathbf{r})$ is the smearing function, which is a $2\times2$
matrix in spinor space. Note that antiquark propagators can be
obtained from quark propagators through
$G_\chi(x,x')=-G_\psi(x',x)^\dag$.  The quantum numbers of the
quarkonium interpolating fields considered in this work are listed in
Table \ref{tab:smear_funcs}. To optimize the overlap with the $1S$ and
$1P$ ground states we use wave functions from a lattice potential model (see Appendix
$D$ of Ref.~\cite{Meinel:2010pv}) in the construction of
$\Gamma(\mathbf{r})$ (as already mentioned in the previous section, the gauge 
configurations are fixed to Coulomb gauge). The heavy-quark mass used in the potential model
is adjusted to match the mass used in the lattice QCD calculation.  We
use $\Gamma(\mathbf{x})$ as the source for the quark propagator and a
point source for the antiquark.  At the sink, the convolution in
Eq.~(\ref{eq:bbarsink}) is performed efficiently using fast Fourier
transforms.

In order to reach the high statistical accuracy needed to extract the
small effects of the isospin charge density, we compute
quarkonium two-point functions for 64 different spatial source
locations (distributed on a cubic sub-lattice with spacing $L/4$) on
each of the source time slices, and average over these source
locations.  Examples of free quarkonium two-point functions with
the smearing technique discussed above are given in
Figs.~\ref{fig:etabUpsiloneffmass} and \ref{fig:heffmass}. Note that
ground-state plateaus are reached already at a distance of $\sim 0.4$
fm in Euclidean time, which demonstrates the efficiency of the
smearing technique used here.

In Table \ref{tab:vac_spec}, we show results for the bottomonium spectrum
in vacuum, from the $16^3\times128$ ensemble at $a_s m=2.75$. To
extract the energies of the $2S$ states, we included additional quarkonium
interpolating fields with the $\phi_{2S}(\mathbf{r})$ smearing in the basis.
The lattice results for the energy splittings are in good agreement with experiment,
confirming the successful tuning of the parameters in the NRQCD action. The remaining
discrepancies are in line with the expected systematic errors (e.g.~discretization errors, missing radiative
and higher-order relativistic corrections in the NRQCD action, and the unphysical pion mass).

\begin{table}[h!]
  \begin{tabular}{llcll}
    \hline\hline
    Name & \hspace{2ex} &  $\mathcal{R}^{PC}$  & \hspace{2ex} & $\Gamma(\mathbf{r})$ \\
    \hline
    $\eta_b$ && $A_1^{-+}$ && $\phi_{1S}(\mathbf{r})$ \\
    $\Upsilon$ && $T_{1}^{--}$ && $\phi_{1S}(\mathbf{r})\:\sigma_j$ \\
    $h_b$ && $T_{1}^{+-}$ && $\phi_{1P}(\mathbf{r}, j)$ \\
    $\chi_{b0}$ && $A_1^{++}$ && $\sum_j \phi_{1P}(\mathbf{r}, j)\:\sigma_j$ \\
    $\chi_{b1}$ && $T_1^{++}$ && $\sum_{k,l} \epsilon_{jkl} \: \phi_{1P}(\mathbf{r}, k) \:\sigma_l$ \\
    $\chi_{b2}$ && $T_2^{++}$ && $\phi_{1P}(\mathbf{r},j)\:\sigma_k + \phi_{1P}(\mathbf{r},k)\:\sigma_j$ \hspace{2ex}(with $j\neq k$)  \\
    \hline\hline
  \end{tabular}
  \caption{\label{tab:smear_funcs}Smearing functions $\Gamma(\mathbf{r})$ used in the quarkonium interpolating fields
  for the given representation of the  cubic group, ${\cal R}$ and values of parity, $P$, and charge-conjugation, $C$. The functions
    $\phi_{1S}({\bf r})$ and $\phi_{1P}(\mathbf{r}, j)$ are eigenfunctions from a lattice potential model.}
\end{table}

\vspace{-4ex}

\begin{table}[h!]
\begin{tabular}{llccc}
\hline\hline
Energy splitting & \hspace{2ex} &  Lattice & \hspace{2ex} & Experiment \\
\hline
$\overline{1P}-\overline{1S}$  && $473.0(3.5)$ && $455.00(92)$ \\
$\overline{2S}-\overline{1S}$  && $569.4(5.0)$ && $572.5(1.4)$ \\
$\Upsilon(1S)-\eta_b(1S)$      && $63.32(42)$  && $62.5(3.6)$ \\
$\Upsilon(2S)-\eta_b(2S)$      && $29.04(59)$  && $24.3(4.5)$ \\
$\chi_{b1}(1P)-\chi_{b0}(1P)$  && $26.31(37)$  && $33.34(66)$ \\
$\chi_{b2}(1P)-\chi_{b1}(1P)$  && $21.12(41)$  && $19.43(57)$ \\
$\overline{1^3P}-h_b(1P)$      && $1.12(22)$   && $0.8(1.1)$ \\
\hline\hline
\end{tabular}
\caption{\label{tab:vac_spec}Bottomonium energy splittings in vacuum, from the $16^3\times 128$ ensemble, for $a_s m=2.75$. All results are in MeV; only
statistical uncertainties are given for the lattice data.
The experimental results for the $\Upsilon(1S)$, $\Upsilon(2S)$, and $\chi_{b\{0,1,2\}}(1P)$ masses are taken from the Particle Data Group \cite{Beringer:1900zz}.
The experimental results for the $h_b(1P)$ and $\eta_b(2S)$ masses are taken from Ref.~\cite{Mizuk:2012pb},
and the experimental $\eta_b(1S)$ mass is the weighted average of the results from Refs.~\cite{Aubert:2008ba, Aubert:2009as, Bonvicini:2009hs, Mizuk:2012pb}, with
a scale factor of 1.9 for the uncertainty (following the PDG procedure for averages).}
\end{table}

\subsection{Correlator ratios for energy shifts}

To investigate the effect of the medium on  quarkonium
propagation, we consider the correlators
\begin{eqnarray}
  C(n; \bbar;t)&=&\langle {\cal O}_\bbar(t){\cal O}_{n\pi^+}(t)\tilde{\cal O}_\bbar^\dagger(0){\cal O}_{n \pi^+}^\dagger(0)\rangle\,,
\end{eqnarray}
where $\langle\ldots\rangle$ denotes path integration via the average
over our ensembles of gauge configurations, and the interpolators
${\cal O}_{n\pi^+}^\dagger$ and ${\cal O}_\bbar^\dagger$ produce the quantum numbers of
$n$-pion and $\bbar$ states  as discussed in the preceding
sub-sections.  States with the combined quantum numbers of the given quarkonium
state ($\bbar=\eta_b,\ \Upsilon,\ h_{b},\ \chi_{b0},\ \chi_{b1},\
\chi_{b2}$) and the $n$-pion system propagate in this correlator and
naturally, the spectrum of this system is different from the sum of
the spectra of $n$ pions and of quarkonium because of interactions. At
Euclidean times where only the ground state of the system is resolved
(after excited states have decayed and before thermal states are
manifest), this correlator will decay exponentially as
\begin{equation}
  C(n; \bbar;t) \longrightarrow  \tilde{Z}_{n;\bbar} \exp(-E_{n;\bbar}t)\,,
\end{equation}
where $E_{n;\bbar}$ is the ground-state energy of the combined system.

To access the change in the quarkonium energy as a function of isospin
density or chemical potential, we further construct the ratios
\begin{eqnarray}
  R(n,\bbar;t) &=& \frac{\langle {\cal O}_\bbar(t){\cal O}_{n\pi^+}(t)\tilde{\cal O}_\bbar^\dagger(0){\cal O}_{n \pi^+}^\dagger(0)\rangle}{\langle {\cal O}_\bbar(t)\tilde{\cal O}_\bbar^\dagger(0)\rangle 
    \langle {\cal O}_{n\pi^+}(t){\cal O}_{n \pi^+}^\dagger(0)\rangle}\,.\hspace*{2mm}
  \label{eq:Rratio}
\end{eqnarray}
Since the
two terms in the denominator decay exponentially at large times as
$\exp(-E_\bbar t)$ and $\exp(-E_{n\pi^+}t)$ respectively, the ratio
will behave as
\begin{equation}
  R(n; \bbar;t) \longrightarrow  Z_{n;\bbar} \exp(-\Delta E_{n;\bbar}t) +\ldots\,,
  \label{eq:expect}
\end{equation}
where $\Delta E_{n;\bbar}=E_{n;\bbar}-E_{n\pi^+} - E_\bbar$ is the
quantity of central interest in our investigation.

\begin{figure}[!th]
  \begin{center}
    \includegraphics[width=\columnwidth]{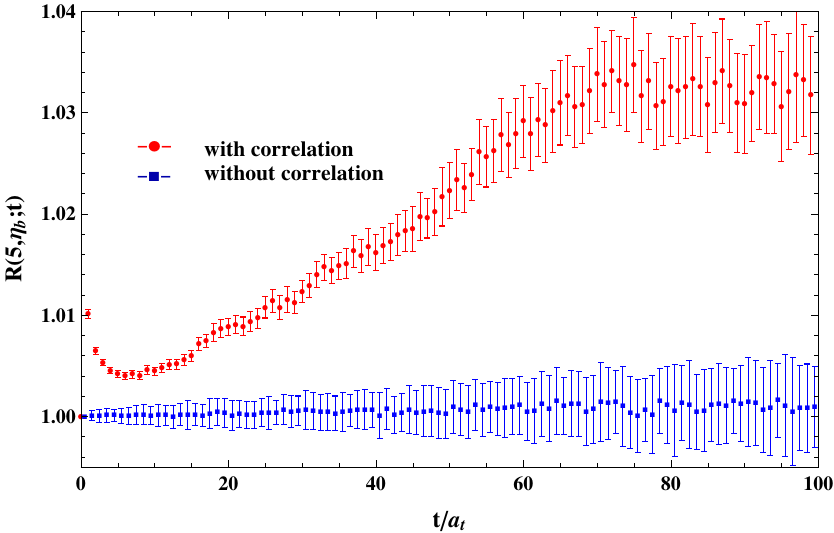}
    \caption{The ratio $R(5,\eta_b;t)$ computed with and without the correct
      correlation between the $\eta_b$ and many-pion system on the $20^3\times256$ ensemble, as discussed in the main text. 
      The time-dependence, which is related to the energy shift through Eq.~(\ref{eq:expect}), only appears when
      correlations are included.}
    \label{fig:without_correlation}
  \end{center}
\end{figure}

As a check of our methods, we constructed ratios in which
we artificially removed the correlations between the $\bbar$ system and
the many-pion state by evaluating $\sum_{c} C_\bbar(c)C_{n
  \pi}(c+\delta c)$, where $C_X(c)$ represents the correlation function for the
quantity $X$ measured on configuration $c$, and $\delta c$ is either a
constant displacement or a random shift. In both cases, the removal
of the correlation eliminates the signal for an energy shift. This is
shown for the $\eta_b$ with $n=5$ in
Fig.~\ref{fig:without_correlation} for random shifts, and the same qualitative 
effect is seen for all choices 
of the density and quarkonium state that are considered.

\section{quarkonium-pion scattering}
\label{sec:UpsilonPi}

The quarkonium state in the presence of a single pion allows us to
study the scattering phase shift of this two-body system using the finite-volume
formalism developed by L\"uscher \cite{Luscher:1986pf,Luscher:1990ux}. The 
$S$-wave quarkonium states we consider
have angular momentum $J=0,1$ and define the total angular momentum of
the entire system since the pion is spin-zero. Since the pion and
$\bbar$ states have different masses, the appropriate generalisation
of the L\"uscher relation to asymmetric systems \cite{Beane:2006gj} is
required. We can define a scattering momentum $p$ through the
relation
\begin{eqnarray}
  \label{eqn_inter_p}
 \sqrt{ (a_s p)^2/\xi^2+a_t^2 M_{\bbar}^2}+\sqrt{ (a_s p)^2/\xi^2+a_t^2M_{\pi}^2} \\
=
a_t\Delta E_{\bbar, \pi} +  a_tM_{\bbar} + a_tM_{\pi} \,,
\nonumber 
\end{eqnarray}
where
 $M_\bbar \equiv M_{\rm kin}^{\bbar}$ is the kinetic mass of the $\bbar$ state. The energy shifts
 $\Delta E_{\bbar, \pi}$ are extracted from fits to the ratios $R(1; \bbar;t)$; see Sec.~\ref{sec:resSwave}
for details of the fitting method and the results for $\Delta E_{\bbar, \pi}$.
 
 The scattering
momentum  then determines the eigenvalue equation
\begin{eqnarray}
  p\cot\delta_{\bbar,\pi}(p)& =& \frac{1}{\pi L} {\bf  S}\left(\frac{p^2 L^2}{4\pi^2}\right), \\
  {\bf  S}(x)&=&\lim_{\Lambda\to\infty}\left[\sum_{{\bf n}\in\mathbb{Z}^3}^{|{\bf n}|<\Lambda}\frac{1}{|{\bf n}|^2-x} -4\pi\Lambda\right]\,,
\end{eqnarray}
that is satisfied by the $\bbar$-$\pi$ scattering phase shift,
$\delta_{\bbar,\pi}(p)$, at the scattering momentum.

\begin{figure}
  \includegraphics[width=0.95\columnwidth]{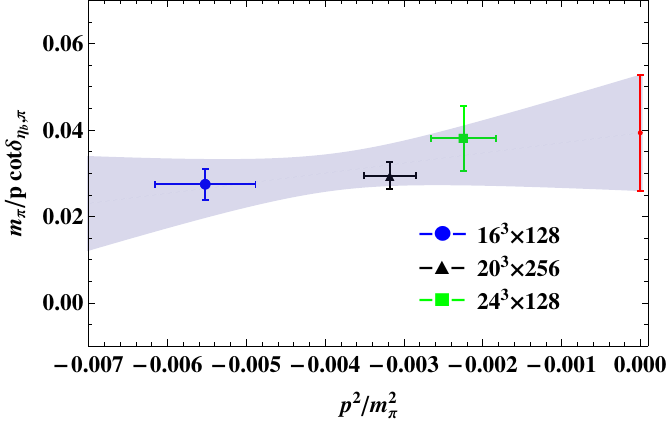}
  \\ \vspace*{4mm}
  \includegraphics[width=0.95\columnwidth]{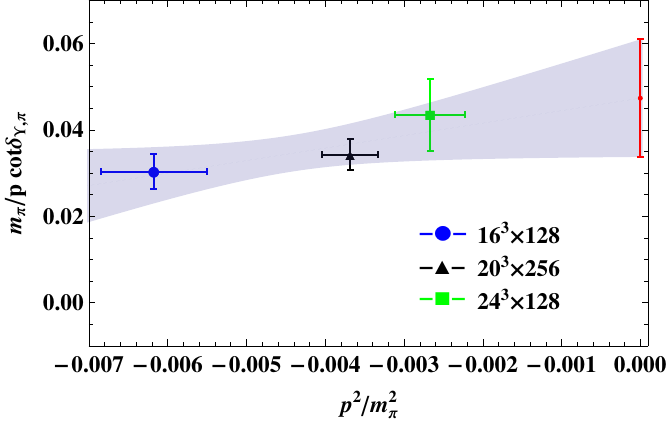}
  \caption{Extracted inverse phase shifts for $\eta_b$-$\pi$ and
    $\Upsilon$-$\pi$ scattering (at $m_\pi\approx 390$ MeV). Fitting the phase shift to $p \cot \delta(p)/m_{\pi}
    = -\frac{1}{m_{\pi}a}+\frac{m_{\pi}r}{2}\frac{p^2}{m_{\pi}^2}$, as
    shown by the shaded band, 
    we can extract the scattering length shown by the point at $p^2/m_\pi^2=0$.}
  \label{fig_scattering_length}
\end{figure}

Since we have three different lattice volumes, we can extract the
phase shift at multiple momenta.  In Figure
\ref{fig_scattering_length}, we show the phase shifts that we extract
for the $\eta_b$-$\pi$ and $\Upsilon$-$\pi$ scattering
channels. These interactions necessarily vanish in the chiral limit as
the quarkonium states are chiral singlet objects \cite{Weinberg:1966kf}. We therefore expect
only small scattering phase shifts at the quark masses considered in
our study. The measured values of the $S$-wave
phase shifts are given in Tables \ref{table:scattering_length_eta} and \ref{table:scattering_length_upsilon},
while for the $P$-wave states we are unable to extract statistically
meaningful results.  Since the measured scattering momenta are small,
it is possible to perform a fit to the effective-range expansion,
\begin{equation}
p \cot \delta(p)/m_{\pi} =
-\frac{1}{m_{\pi}a}+\frac{m_{\pi}r}{2}\frac{p^2}{m_{\pi}^2} +\ldots\,,
\end{equation}
 to extract the scattering length and effective range for these
interactions. This extrapolation is shown in 
Fig.~\ref{fig_scattering_length} and results in $m_{\pi}a_{\eta_b,\pi}
= 0.039(13)$ and
$m_\pi r_{\eta_b, \pi} = 4.7(3.7)$
 for the $\eta_b$ state,
and $m_{\pi}a_{\Upsilon,\pi} = 0.047(14)$ and $m_\pi r_{\Upsilon, \pi}= 5.8(3.3)$
 in the case of the $\Upsilon$, both channels corresponding to a weak
 attractive interaction.

The pion-quarkonium scattering length depends approximately quadratically on the pion mass
\cite{Yokokawa:2006td, Liu:2008rza, Liu:2009af}, and hence we can estimate the scattering
length at the physical pion mass as
\begin{equation}
 a_{\bbar,\pi}^{(\rm phys.)} \approx (m_\pi^{(\rm phys.)}/m_\pi)^2 \: a_{\bbar,\pi}, \label{eq:SLinterp}
\end{equation}
where $a_{\bbar,\pi}$ is our lattice result for the scattering length
at $m_\pi=390$ MeV. This gives
\begin{equation}
 a_{\eta_b,\pi}^{(\rm phys.)} = 0.0025(8)(6)\:\:{\rm fm},\hspace{2ex} a_{\Upsilon,\pi}^{(\rm phys.)} = 0.0030(9)(7)\:\:{\rm fm}, \label{eq:SLphys}
\end{equation}
where the first uncertainty is statistical and the second uncertainty corresponds to missing higher-order corrections to Eq.~(\ref{eq:SLinterp}),
which we estimate to be smaller than the leading-order term by a factor of $m_\pi/(4\pi f_\pi)\approx0.24$.
Related lattice QCD calculations of charmonium-pion scattering lengths were reported in Refs.~\cite{Yokokawa:2006td, Liu:2008rza, Liu:2009af},
and model-dependent studies of quarkonium-pion interactions can be found in Refs.~\cite{Peskin:1979va, Bhanot:1979vb, Fujii:1999xn, Liu:2012dv}.
In general, similarly small attractive interactions were found there.
 
\begin{table}[ht]
  \caption{The $\eta_b$-$\pi$ phase shifts (at $m_\pi\approx 390$ MeV) extracted using the L\"uscher method.}
  \begin{center}
    \begin{tabular}{c c  c  c c  }
    \hline\hline
    $N_s^3 \times N_t$ & $p^2/m_{\pi}^2$ & ${( p \cot\delta(p))^{-1}} [{\rm fm}]$ & $m_\pi/(p \cot\delta(p))$ 
     \\
\hline 
    $16^3 \times 128$ &$-0.0055(6) $ & $0.0138(18)$ & $0.0274(36)$ \\
    $20^3 \times 256$ & $-0.0032(3)$ & $0.0148(15) $ & $0.0294(31)$ \\
    $24^3 \times 128$ &$-0.0022(4) $ & $0.0192(38) $  & $0.0381(75)$ \\
    \hline\hline
  \end{tabular}
  \vspace{-4ex}
\end{center}
    \label{table:scattering_length_eta}
\end{table}

\begin{table}[ht]
  \caption{The $\Upsilon$-$\pi$ phase shifts (at $m_\pi\approx 390$ MeV) extracted using the L\"uscher method.}
  \begin{center}
  \begin{tabular}{c   cc  c c }
    \hline\hline
    $N_s^3 \times N_t$ & $p^2/m_{\pi}^2$ & ${( p \cot\delta(p))^{-1}} [{\rm fm}]$ & $m_\pi/(p \cot\delta(p))$ 
    \\
    \hline
    $16^3 \times 128$  &$-0.0062(7) $ & $0.0153(20)$ & $0.0303(40)$ \\
    $20^3 \times 256$  &$-0.0037(4) $ & $0.0172(18)$ & $0.0341(36)$ \\
    $24^3 \times 128$  &$-0.0027(4) $ & $0.0220(42)$ & $0.0435(83)$ \\
    \hline\hline
  \end{tabular}
 \vspace{-4ex}
\end{center}
  \label{table:scattering_length_upsilon}
\end{table}

\section{Isospin density dependence of quarkonium}
\label{sec:res}

For larger isospin charge, we interpret the system of pions in terms
of a medium of varying isospin charge density once the ground state is
reached. In the correlators $C(n;\bbar;t)$, the quarkonium state
exists in this medium, interacting with it. We consider first the
$S$-wave quarkonium states as they are statistically better resolved
than $P$-wave states.
\vspace{-4ex}

\subsection{ $S$-wave states}

\label{sec:resSwave}

\begin{figure*}
  \begin{center}
    \includegraphics[width=0.3\textwidth]{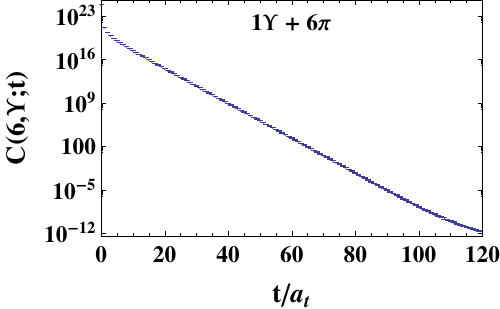}
    \quad
    \includegraphics[width=0.3\textwidth]{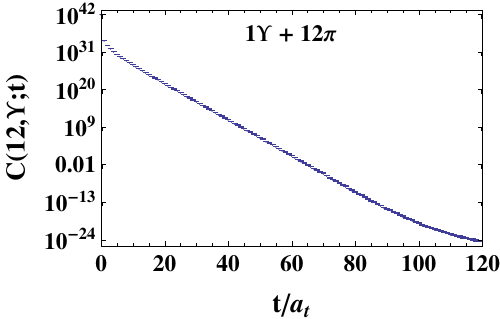}
    \quad
   \includegraphics[width=0.3\textwidth]{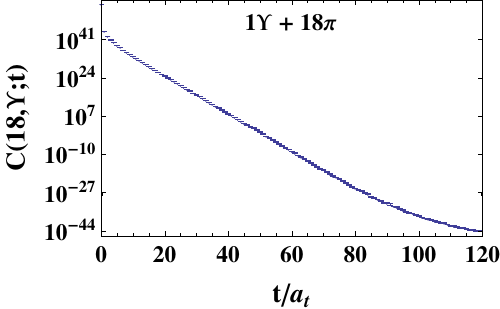}
   \\ 
     \includegraphics[width=0.3\textwidth]{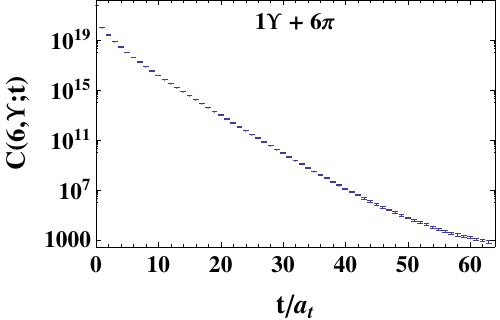} 
     \quad
     \includegraphics[width=0.3\textwidth]{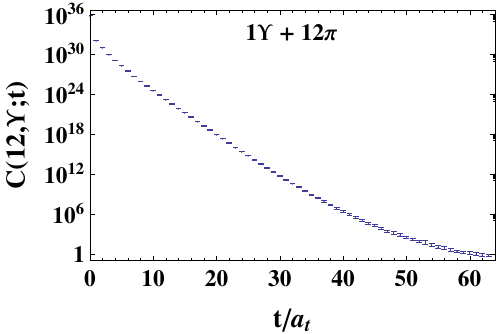}
    \quad
    \includegraphics[width=0.3\textwidth]{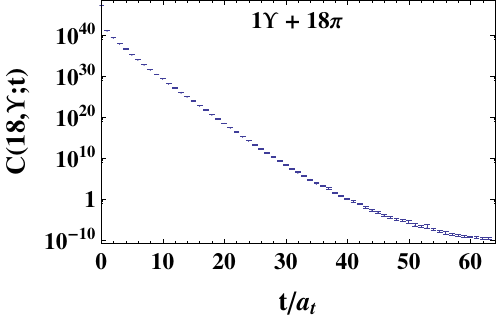}
    \caption{The correlators for the $\Upsilon$ in a medium
      corresponding to isospin charge $n$ for
      $n=6$, 12, and 18 are shown. Data are presented for $a_s m=2.75$ on
      the $20^3\times256$  (upper) and $16^3\times128$
      (lower) ensembles. Correlators for the $\eta_b$ in medium behave similarly.
 }
    \label{fig:correlator_numerator}
  \end{center}
\end{figure*}

The correlators $C(n,\bbar,t)$ are shown in
Fig.~\ref{fig:correlator_numerator} for $\bbar=\Upsilon$ 
at representative values of the isospin
charge and for  $a_s m=2.75$ on the $20^3\times256$ and $16^3\times128$
ensembles.  The in-medium correlators on the $20^3\times256$ ensemble 
exhibit a long region of Euclidean
time in which they decay as a single exponential.  This region
overlaps with the regions in which the multi-pion correlators and the
individual quarkonium correlators are saturated by their respective
ground states. This gives us confidence that by considering the
correlator ratios of Eq.~(\ref{eq:Rratio}) we can legitimately extract
the quarkonium energy shifts in medium. On the ensembles with $T=128$, 
 thermal contamination is more significant and restricts the range of
 useful time-slices, particularly for large isospin
charge.

\begin{figure*}
  \begin{center}
    \includegraphics[width=0.3\textwidth]{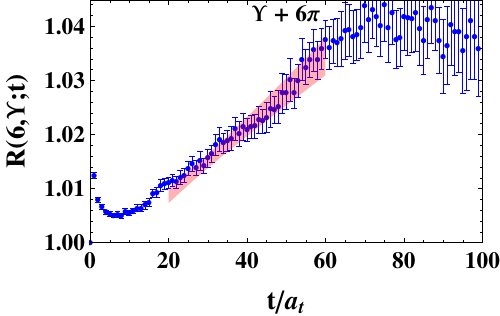}
    \quad
    \includegraphics[width=0.3\textwidth]{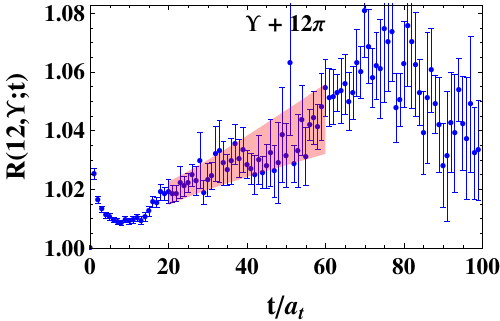}
    \quad
    \includegraphics[width=0.3\textwidth]{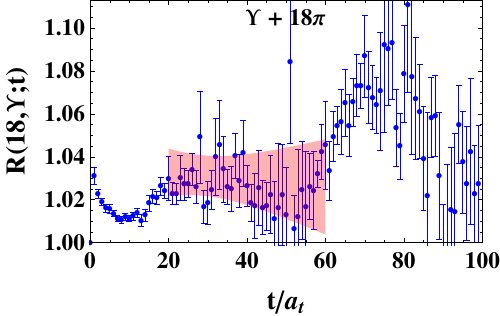} 
    \\
    \includegraphics[width=0.3\textwidth]{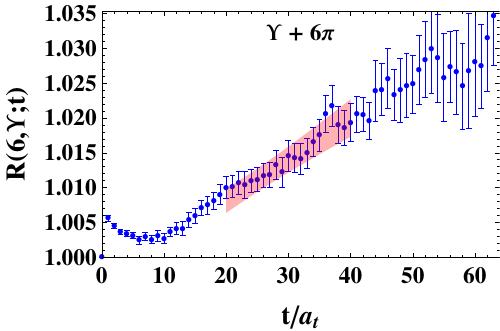}
    \quad
    \includegraphics[width=0.3\textwidth]{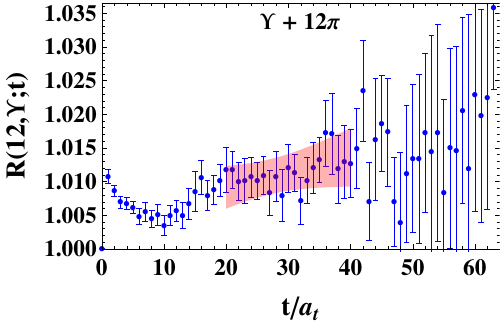}
    \quad
    \includegraphics[width=0.3\textwidth]{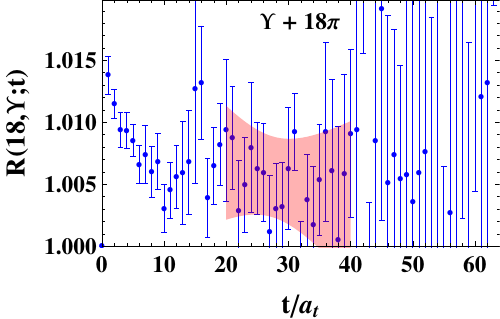}
    \caption{The correlator ratios for the $\Upsilon$ in a medium
      corresponding to isospin charges $n= 6$, 12, 18.
      The shaded bands show the statistical uncertainties of fits of
      the form given in Eq.~(\ref{eq:expect}).
      Data are shown for $a_s m=2.75$
      on the $20^3\times256$ (upper) and $16^3\times128$ (lower) ensembles.}
    \label{fig:correlator_ratios}
  \end{center}
\end{figure*}

\begin{figure*}
  \begin{center}
    \includegraphics[width=0.3\textwidth]{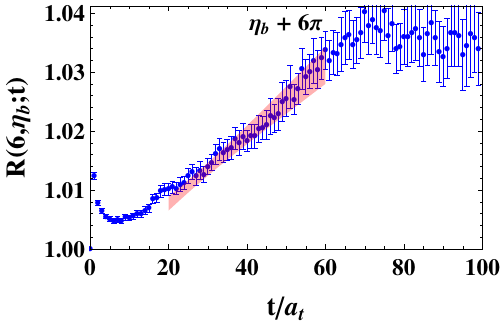}
    \quad
    \includegraphics[width=0.3\textwidth]{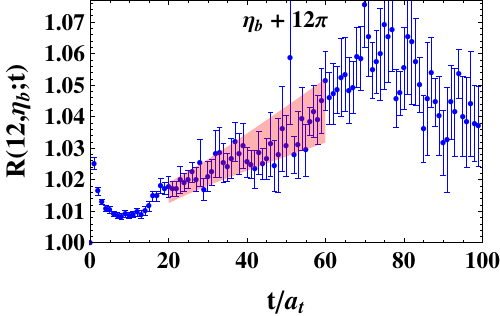} 
    \quad
    \includegraphics[width=0.3\textwidth]{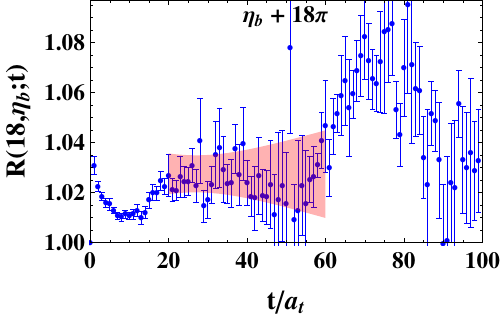} 
    \\    
    \includegraphics[width=0.3\textwidth]{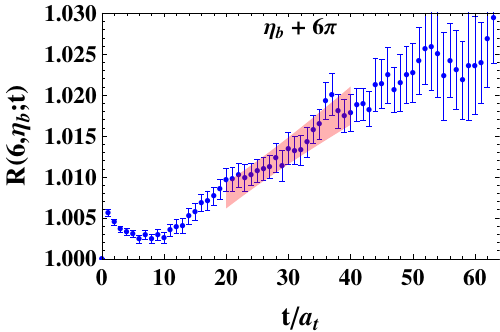}
    \quad
    \includegraphics[width=0.3\textwidth]{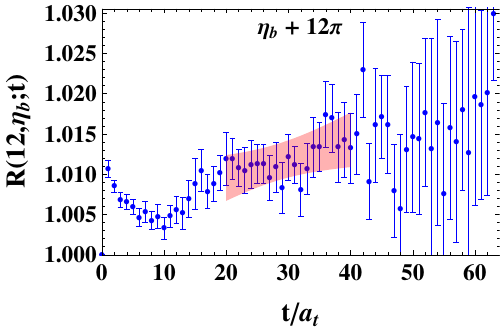}
    \quad
    \includegraphics[width=0.3\textwidth]{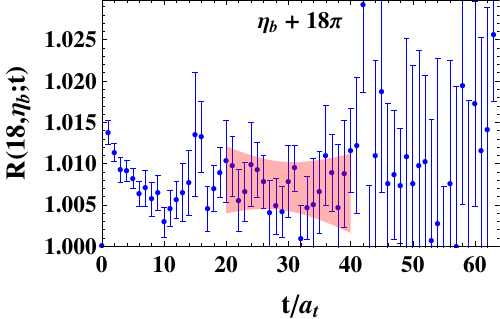}
    \caption{The correlator ratios for the $\eta_b$ in a medium
      corresponding to isospin charges $n=6$, 12, 18.
      The shaded bands show the statistical uncertainties of fits of
      the form given in Eq.~(\ref{eq:expect}).
      Data are shown for $a_s m=2.75$ on the
      $20^3\times256$ (upper) and $16^3\times128$ (lower) ensembles.}
    \label{fig:correlator_ratios_eta}
  \end{center}
\end{figure*}

\begin{table*}
  \begin{tabular}{cc|cccccc|ccccccc}
    \hline\hline
    $n$ & \hspace{2ex} &   & \hspace{2ex} & $\Delta E_{n;\eta_b}$ & \hspace{2ex} & & \hspace{2ex} &   & \hspace{2ex} & $\Delta E_{n;\Upsilon}$& \hspace{2ex} &  \\
    & \hspace{2ex} &  $16^3\times128$  & \hspace{2ex} & $20^3\times256$& \hspace{2ex} & $24^3\times128$  & \hspace{2ex} &  $16^3\times128$  & \hspace{2ex} & $20^3\times256$& \hspace{2ex} & $24^3\times128$ \\
    \hline
 
$1$&\hspace{2ex} & $-1.12(11)(08)$&\hspace{2ex} &  $-0.62(06)(02)$&\hspace{2ex} &  $-0.46(06)(06)$&\hspace{2ex} &  $-1.23(12)(09)$&\hspace{2ex} &  $-0.72(07)(03)$&\hspace{2ex} &  $-0.53(07)(06)$\\ 
$2$&\hspace{2ex} & $-1.95(21)(14)$&\hspace{2ex} &  $-1.20(12)(06)$&\hspace{2ex} &  $-0.89(13)(10)$&\hspace{2ex} &  $-2.15(23)(15)$&\hspace{2ex} &  $-1.38(14)(07)$&\hspace{2ex} &  $-1.01(15)(11)$\\ 
$3$&\hspace{2ex} & $-2.51(30)(18)$&\hspace{2ex} &  $-1.74(19)(12)$&\hspace{2ex} &  $-1.26(21)(13)$&\hspace{2ex} &  $-2.75(34)(20)$&\hspace{2ex} &  $-1.99(22)(13)$&\hspace{2ex} &  $-1.44(23)(14)$\\ 
$4$&\hspace{2ex} & $-2.83(40)(21)$&\hspace{2ex} &  $-2.25(28)(19)$&\hspace{2ex} &  $-1.57(29)(14)$&\hspace{2ex} &  $-3.08(45)(23)$&\hspace{2ex} &  $-2.54(31)(21)$&\hspace{2ex} &  $-1.80(31)(16)$\\ 
$5$&\hspace{2ex} & $-2.97(51)(26)$&\hspace{2ex} &  $-2.73(37)(28)$&\hspace{2ex} &  $-1.81(37)(16)$&\hspace{2ex} &  $-3.23(58)(29)$&\hspace{2ex} &  $-3.04(40)(28)$&\hspace{2ex} &  $-2.08(41)(18)$\\ 
$6$&\hspace{2ex} & $-2.99(61)(31)$&\hspace{2ex} &  $-3.17(47)(37)$&\hspace{2ex} &  $-1.97(47)(18)$&\hspace{2ex} &  $-3.23(70)(37)$&\hspace{2ex} &  $-3.47(51)(36)$&\hspace{2ex} &  $-2.27(51)(20)$\\ 
$7$&\hspace{2ex} & $-2.89(71)(37)$&\hspace{2ex} &  $-3.53(58)(46)$&\hspace{2ex} &  $-2.05(58)(22)$&\hspace{2ex} &  $-3.10(81)(45)$&\hspace{2ex} &  $-3.81(61)(45)$&\hspace{2ex} &  $-2.37(63)(24)$\\ 
$8$&\hspace{2ex} & $-2.69(81)(41)$&\hspace{2ex} &  $-3.80(70)(54)$&\hspace{2ex} &  $-2.05(71)(29)$&\hspace{2ex} &  $-2.86(92)(51)$&\hspace{2ex} &  $-4.03(73)(53)$&\hspace{2ex} &  $-2.38(77)(31)$\\ 
$9$&\hspace{2ex} & $-2.40(89)(44)$&\hspace{2ex} &  $-3.95(83)(62)$&\hspace{2ex} &  $-1.97(86)(38)$&\hspace{2ex} &  $-2.5(1.0)(0.6)$&\hspace{2ex} &  $-4.12(86)(62)$&\hspace{2ex} &  $-2.31(93)(41)$\\ 
$10$&\hspace{2ex} & $-2.05(97)(47)$&\hspace{2ex} &  $-3.95(96)(72)$&\hspace{2ex} &  $-1.8(1.0)(0.5)$&\hspace{2ex} &  $-2.1(1.2)(0.6)$&\hspace{2ex} &  $-4.1(1.0)(0.7)$&\hspace{2ex} &  $-2.2(1.1)(0.5)$\\ 
$11$&\hspace{2ex} & $-1.7(1.1)(0.5)$&\hspace{2ex} &  $-3.8(1.1)(0.8)$&\hspace{2ex} &  $-1.6(1.2)(0.6)$&\hspace{2ex} &  $-1.7(1.3)(0.7)$&\hspace{2ex} &  $-3.8(1.2)(0.9)$&\hspace{2ex} &  $-1.9(1.3)(0.7)$\\ 
$12$&\hspace{2ex} & $-1.3(1.2)(0.7)$&\hspace{2ex} &  $-3.5(1.2)(1.0)$&\hspace{2ex} &  $-1.3(1.4)(0.8)$&\hspace{2ex} &  $-1.2(1.4)(0.8)$&\hspace{2ex} &  $-3.4(1.4)(1.1)$&\hspace{2ex} &  $-1.6(1.5)(0.8)$\\ 
$13$&\hspace{2ex} & $-0.9(1.3)(0.8)$&\hspace{2ex} &  $-3.1(1.4)(1.2)$&\hspace{2ex} &  $-1.0(1.6)(1.0)$&\hspace{2ex} &  $-0.8(1.6)(1.0)$&\hspace{2ex} &  $-2.8(1.7)(1.3)$&\hspace{2ex} &  $-1.3(1.8)(1.0)$\\ 
$14$&\hspace{2ex} & $-0.6(1.4)(1.0)$&\hspace{2ex} &  $-2.5(1.6)(1.5)$&\hspace{2ex} &  $-0.6(1.9)(1.1)$&\hspace{2ex} &  $-0.4(1.8)(1.2)$&\hspace{2ex} &  $-2.1(2.0)(1.6)$&\hspace{2ex} &  $-1.0(2.0)(1.2)$\\ 
$15$&\hspace{2ex} & $-0.3(1.5)(1.3)$&\hspace{2ex} &  $-1.9(1.8)(1.8)$&\hspace{2ex} &  $-0.3(2.1)(1.3)$&\hspace{2ex} &  $-0.0(2.0)(1.4)$&\hspace{2ex} &  $-1.3(2.4)(1.9)$&\hspace{2ex} &  $-0.6(2.3)(1.4)$\\ 
$16$&\hspace{2ex} & $-0.0(1.6)(1.5)$&\hspace{2ex} &  $-1.2(2.1)(2.1)$&\hspace{2ex} &  $0.1(2.4)(1.5)$&\hspace{2ex} &  $0.3(2.1)(1.7)$&\hspace{2ex} &  $-0.5(2.8)(2.2)$&\hspace{2ex} &  $-0.2(2.6)(1.6)$\\ 
$17$&\hspace{2ex} & $0.2(1.7)(1.8)$&\hspace{2ex} &  $-0.6(2.3)(2.4)$&\hspace{2ex} &  $0.4(2.7)(1.7)$&\hspace{2ex} &  $0.6(2.3)(1.9)$&\hspace{2ex} &  $0.2(3.1)(2.4)$&\hspace{2ex} &  $0.2(2.9)(1.8)$\\ 
$18$&\hspace{2ex} &  $0.5(1.8)(2.0)$&\hspace{2ex} &  $0.0(2.6)(2.7)$&\hspace{2ex} &  $0.7(3.0)(1.8)$&\hspace{2ex} &  $0.8(2.4)(2.2)$&\hspace{2ex} &  $0.9(3.5)(2.6)$&\hspace{2ex} &  $0.5(3.2)(2.0)$\\ 
$19$&\hspace{2ex} &  $0.7(1.9)(2.3)$&\hspace{2ex} &  $0.6(2.8)(2.9)$&\hspace{2ex} &  $0.9(3.3)(2.0)$&\hspace{2ex} &  $1.1(2.5)(2.4)$&\hspace{2ex} &  $1.5(3.8)(2.8)$&\hspace{2ex} &  $0.8(3.5)(2.2)$\\ 
$20$&\hspace{2ex} &  $0.9(1.9)(2.5)$&\hspace{2ex} &  $1.1(3.0)(3.0)$&\hspace{2ex} &  $1.1(3.6)(2.2)$&\hspace{2ex} &  $1.3(2.5)(2.6)$&\hspace{2ex} &  $2.1(4.0)(2.9)$&\hspace{2ex} &  $1.0(3.8)(2.4)$\\
     \hline\hline
  \end{tabular}
  \caption{\label{tab:Sfits}Energy shifts in MeV from fits to the $S$-wave correlator ratios on
    the various ensembles, for $a_s m=2.75$. For each combination, we
    report: the mean and the statistical and systematic uncertainties.
   }
\end{table*}

\begin{figure*}
  \begin{center}
    \includegraphics[width=0.49\linewidth]{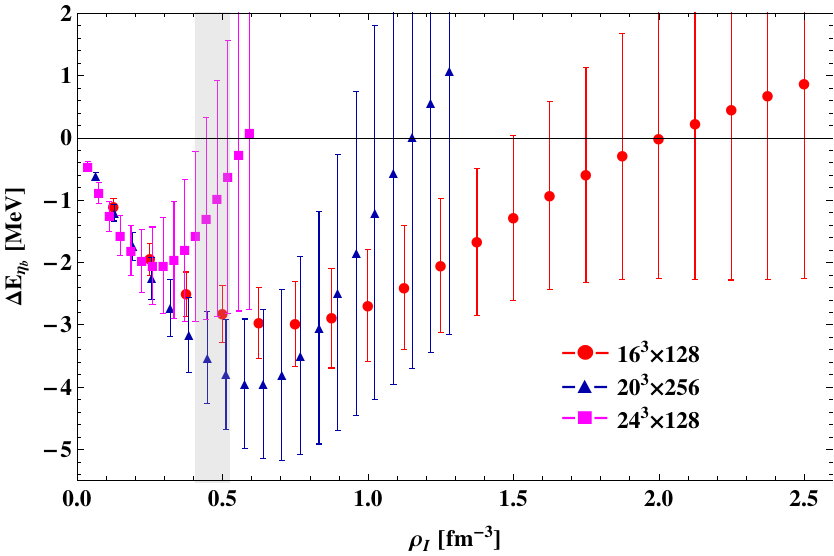}
    \hfill
    \includegraphics[width=0.49\linewidth]{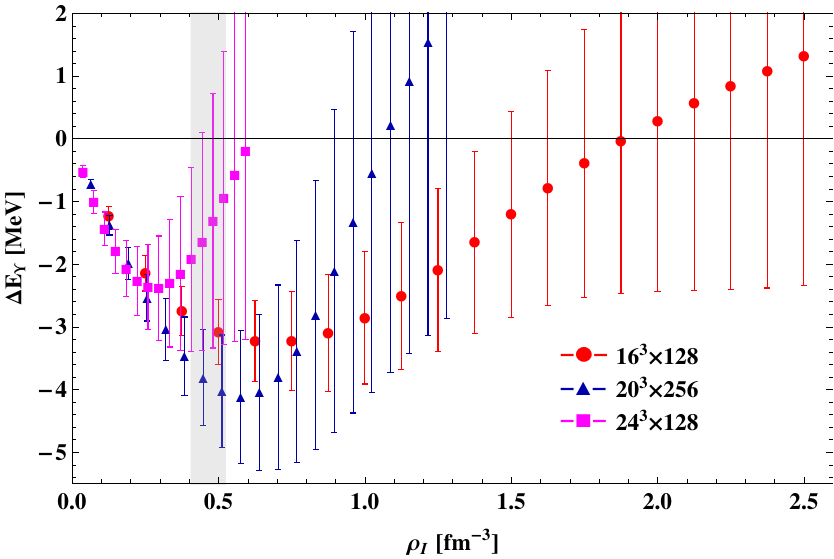}
    \caption{The dependence of the energy shift on the isospin
      charge density is shown for the three lattice volumes for
      the $\eta_b$ (left panel) and $\Upsilon$ (right panel). The results are for $a_s m=2.75$. The
      shaded vertical band in each plot shows the region where there
      is a peak in the ratio of the pionic energy density to the
      Stefan-Boltzmann expectation (see Fig.~22 of Ref.~\protect\cite{Detmold:2012wc}).   }
    \label{fig:deltaE_vs_mu}
  \end{center}
\end{figure*}

\begin{figure*}
  \begin{center}
    \includegraphics[width=0.49\linewidth]{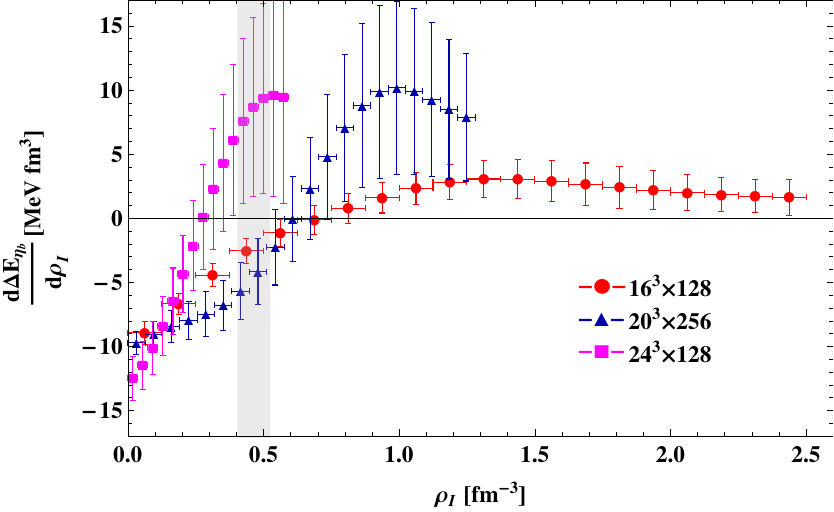}
    \hfill
    \includegraphics[width=0.49\linewidth]{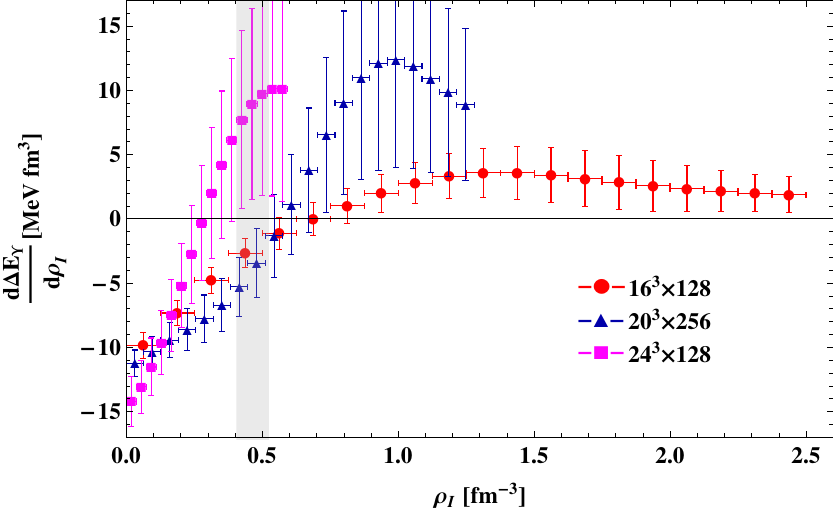}
    \caption{The slope $\mathrm{d}(\Delta E)/\mathrm{d}\rho_I$ of the $\eta_b$ energy shift (left panel) and $\Upsilon$ energy shift (right panel),
    approximated using correlated finite differences. The data sets and shaded bands are as described in Fig.~\ref{fig:deltaE_vs_mu}.}
    \label{fig:d_deltaE_vs_mu}
  \end{center}
\end{figure*}

\begin{figure}
  \begin{center}
    \includegraphics[width=\columnwidth]{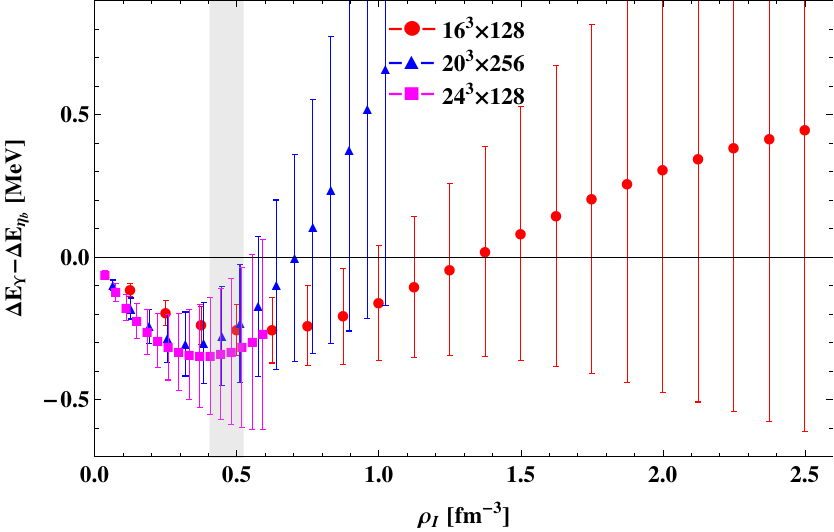}
    \caption{Isospin density dependence of the shift of the
      $S$-wave hyperfine splitting between the  $\Upsilon$ and $\eta_b$
      states in medium. The results are for $a_s m=2.75$.}
    \label{fig:hypersplit_vs_mu}
  \end{center}
\end{figure}

The correlator ratios, $R(n,\bbar;t)$, discussed above, are shown for
both $\Upsilon$ and $\eta_b$ at a heavy quark mass $a_s m=2.75$
on the $20^3\times256$ ensemble for a range of different isospin
charges, $n= 6$, 12, and 18, in Figs.~\ref{fig:correlator_ratios}
and \ref{fig:correlator_ratios_eta} along with fits to time dependence
using Eq.~(\ref{eq:expect}). Fits are performed over a range of times
where both the individual multi-pion correlation functions and
quarkonium correlation functions exhibit ground-state saturation and
are free from thermal (backward propagating) state contamination. This
is ensured by choosing the central fit range $[t_{\rm min},t_{\rm
  max}]$ such that a fit over the range $[t_{\rm min}-5, t_{\rm
  max}+5]$ has an acceptable quality of fit. On the $20^3\times256$
ensemble, we choose $t_{\rm min}=20$ and $t_{\rm max}=60$, beyond
which thermal contributions are apparent. For the ensembles with $T=128$, we choose $t_{\rm
    max}=40$. Statistical
uncertainties are estimated using the bootstrap procedure. To estimate
the systematic uncertainties of the fits, we calculate the standard deviation
between the three energies extracted from fits with the ranges $[t_{\rm min}-5, t_{\rm
  max}-5]$, $[t_{\rm min}, t_{\rm max}]$, and $[t_{\rm min}+5, t_{\rm
  max}+5]$ on each bootstrap sample. The systematic uncertainty is then obtained as
  the average of this standard deviation over the bootstrap samples.
  On the $20^3\times256$ ensemble, the correlator ratios show some long-range oscillations at large $t$,
  and there we use the three ranges $[t_{\rm min}-5, t_{\rm  max}-20]$,
  $[t_{\rm min}, t_{\rm max}]$, and $[t_{\rm min}+5, t_{\rm  max}+20]$ to estimate the systematic fitting
  uncertainty.

  The extracted energy shifts and uncertainties are shown in
Table \ref{tab:Sfits}.  For larger values of $n$, the energy shifts
become noisier and we  limit our analysis to the range of isospin densities 
where a successful fit could be performed for a given ensemble.

To summarise the analysis of the correlator ratios for the S-wave
quarkonium states, Fig.~\ref{fig:deltaE_vs_mu} shows the isospin
density dependence of the energy shifts, $\Delta E_{n;\bar{b}b}$, for both the $\Upsilon$ and $\eta_b$
channels. Figure \ref{fig:d_deltaE_vs_mu} additionally shows the derivative
$\mathrm{d}(\Delta E)/\mathrm{d}\rho_I$, approximated by the finite difference $(\Delta E_{n;\bar{b}b}-\Delta E_{(n-1);\bar{b}b}) L^3$,
taking into account the strong correlations between the energies at different $n$.
Results are presented for the ranges of isospin
charge density where a statistically meaningful extraction of the
energy shift can be made. As can be seen in Fig.~\ref{fig:deltaE_vs_mu}, there is a 
significant negative energy shift for much of the range of isospin density
 that we have investigated. The magnitude of this shift first increases as the
isospin density is increased, before flattening off at a value of about
3 MeV and possibly decreasing for large $\rho_I$, albeit with 
increasing uncertainty. A consistent picture is found from the derivatives
shown in Fig.~\ref{fig:d_deltaE_vs_mu}.  It
is interesting to note that the saturation occurs at the point at
which a
marked change in the energy density of the many-pion system was observed in Ref.~ \cite{Detmold:2012wc}, 
and is
likely caused by the changing nature of the screening medium at this
point.  The increase of the energy shift at low densities is in line with the expectations of 
the potential model discussed earlier, but the energy shift is numerically larger than in the model (note
that the potential model was based on lattice results for the screening of the static potential at $m_\pi\sim320$ MeV \cite{Detmold:2008bw},
whereas the present NRQCD calculations were done with $m_\pi \sim 390$ MeV).
The saturation effect was not predicted by the model; since the model was developed using the measured shifts in the
potential in the low density region, so this is not surprising.

We have performed these calculations for all three ensembles
of configurations but have only been able to access a limited range of
densities with the current statistical precision. The $16^3\times128$ ensemble provides the largest density range.
The results from all of the ensembles are consistent in the region in which they
overlap. The zero-crossings of the derivatives
in Fig.~\ref{fig:d_deltaE_vs_mu} on the $24^3\times128$ ensemble are at a slightly lower
isospin density than on the other two ensembles, but the difference is not significant.

We also consider the shifts in the  splitting between the 
$\eta_b$ and $\Upsilon$ energies in medium as a function of the density. 
We extract these shifts  by calculating the correlated differences
between the individual energies using the bootstrap method.
A summary of the isospin
charge dependence of this splitting is shown in
Fig.\ref{fig:hypersplit_vs_mu}. It can be seen that the $\Upsilon$
energy is shifted slightly more than the $\eta_b$ energy by the presence of the medium.

\FloatBarrier

\subsection{ $P$-wave states}

We  also analyse the lowest-energy $P$-wave quarkonium states, $h_b$,
$\chi_{b0}$, $\chi_{b1}$ and $\chi_{b2}$, in medium. We find that we
cannot resolve differences between the medium effects for these
different states and so consider a spin average of their energies. In
order to extract the spin-averaged in-medium energy shift
\begin{eqnarray}
 \nonumber \Delta E_{n;\overline{1P}} &=& \sfrac{3}{12}\Delta E_{n;h_b}+\sfrac{1}{12} \Delta E_{n;\chi_{b0}} \\
     &&  +\sfrac{3}{12}\Delta E_{n;\chi_{b1}}+\sfrac{5}{12} \Delta E_{n;\chi_{b2}},
\end{eqnarray}
we construct the following
product of fractional powers of the individual ratios,
\begin{eqnarray}
\nonumber R(n,\overline{1P};t) &=& R(n,h_b;t)^{\frac{3}{12}} \:\: R(n,\chi_{b0};t)^{\frac{1}{12}}\\
 && \times R(n,\chi_{b1};t)^{\frac{3}{12}} \:\: R(n,\chi_{b2};t)^{\frac{5}{12}},
\end{eqnarray}
which at large $t$ will behave as
\begin{equation}
 R(n,\overline{1P};t) \longrightarrow Z_{n;h_b}^{\frac{3}{12}} Z_{n;\chi_{b0}}^{\frac{1}{12}} Z_{n;\chi_{b1}}^{\frac{3}{12}} Z_{n;\chi_{b2}}^{\frac{5}{12}} \:\exp (-\Delta E_{n;\overline{1P}} t).
\end{equation}
We also consider the analogous $S$-wave spin-average combination
\begin{equation}
 R(n,\overline{1S};t) = R(n,\eta_b;t)^{\frac14} \:\: R(n,\Upsilon;t)^{\frac34}.
\end{equation}
Since the $P$-wave quarkonium correlators are themselves statistically noisier than the
$S$-wave correlators (see Figs.~\ref{fig:etabUpsiloneffmass} and 
\ref{fig:heffmass}), the precision with which we can extract the
$P$-wave energy shifts is
reduced.

\begin{figure*}
  \begin{center}
    \includegraphics[width=0.3\textwidth]{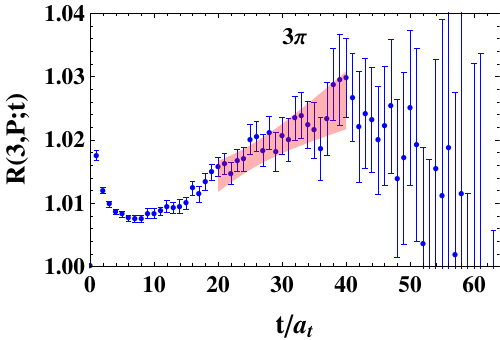}
  \quad
    \includegraphics[width=0.3\textwidth]{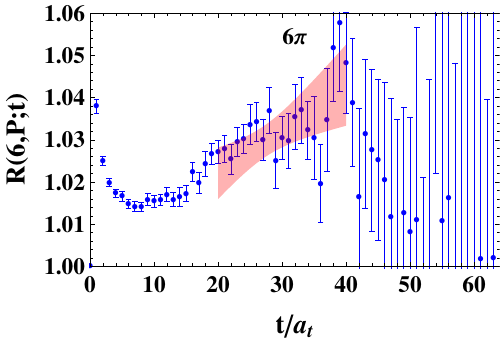}
    \quad
    \includegraphics[width=0.3\textwidth]{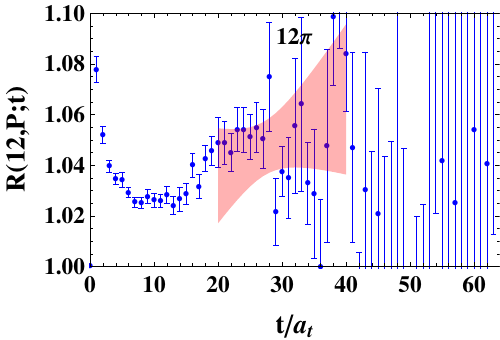} 
\\
    \includegraphics[width=0.3\textwidth]{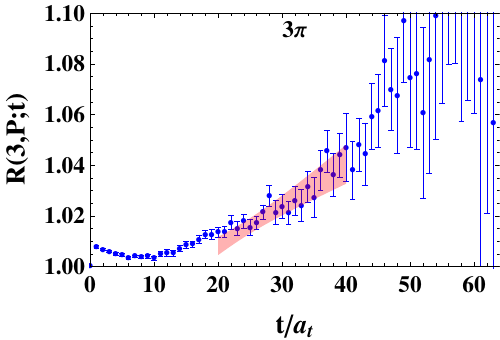}
    \quad
    \includegraphics[width=0.3\textwidth]{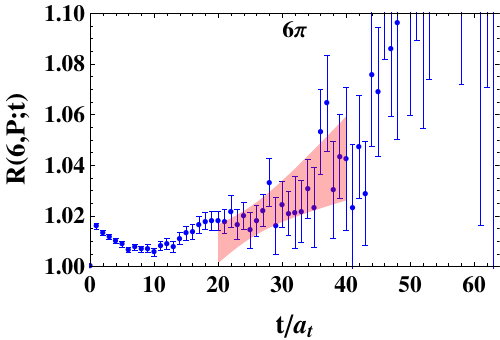}
    \quad
    \includegraphics[width=0.3\textwidth]{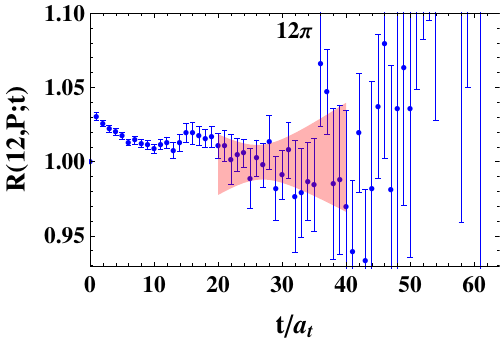}
    \caption{The correlators ratios 
    corresponding to the spin-averaged $P$-wave energy in a
      medium corresponding to isospin charges $n=3$, 6, and 12. Data are shown for $a_s m=2.75$ on the
      $20^3\times256$ (upper) and $16^3\times128$ (lower) ensembles.}
    \label{fig:correlaor_ratios_pwave}
  \end{center}
\end{figure*}
 
\begin{figure}
  \begin{center}
    \includegraphics[width=0.95\columnwidth]{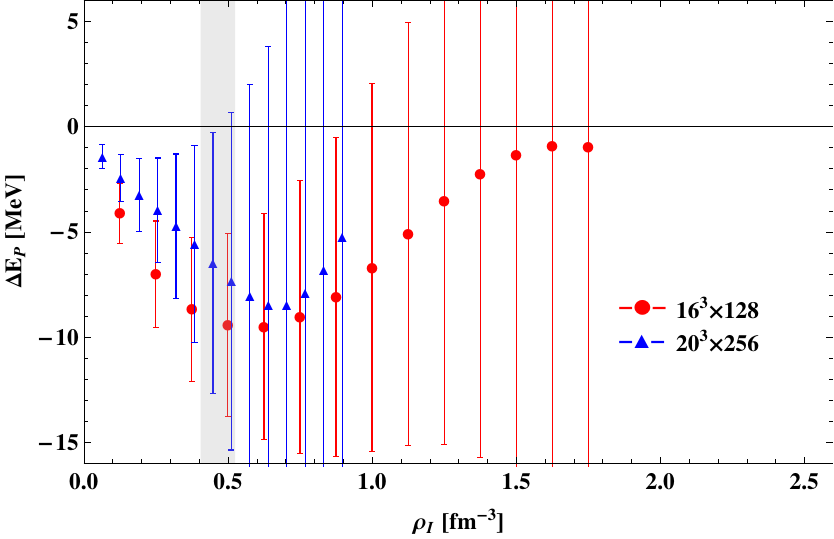}\\
    \vspace*{5mm}
    \includegraphics[width=0.95\columnwidth]{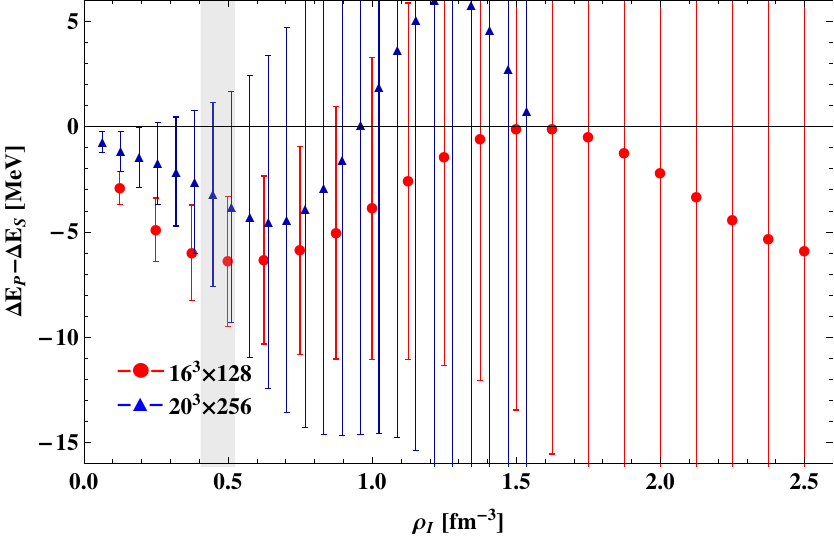}
    \caption{Upper panel: the shift in the spin-averaged $1P$ energy as a function of the isospin
      charge density. Lower panel: the shift of the spin-averaged $\overline{1P}-\overline{1S}$
      splitting. The vertical band shows
      the isospin density at which the pionic energy density is peaked
      relative to the Stefan-Boltzmann expectation.
      The results are for $a_s m=2.75$.}
    \label{fig:shifts_pwave}
  \end{center}
\end{figure}

Figure \ref{fig:correlaor_ratios_pwave} shows representative
correlator ratios $R(n,\overline{1P};t)$,
and Fig.~\ref{fig:shifts_pwave} summarises the extracted
energy shifts. Here we only show results from the $16^3\times 128$ and $20^3\times 256$ ensembles,
because the $P$-wave results on the $24^3\times 128$ ensemble were too noisy.
The potential model expectation is that the $P$-wave shift will be larger than the
$S$-wave shift, and our lattice results confirm
the expectation. In the lower panel of Fig.~\ref{fig:shifts_pwave} we show the correlated differences
between the spin-averaged $P$-wave and $S$-wave energy shifts.

\subsection{Heavy-quark mass dependence}

\begin{figure}[!t]
  \begin{center}
    \includegraphics[width=0.95\columnwidth]{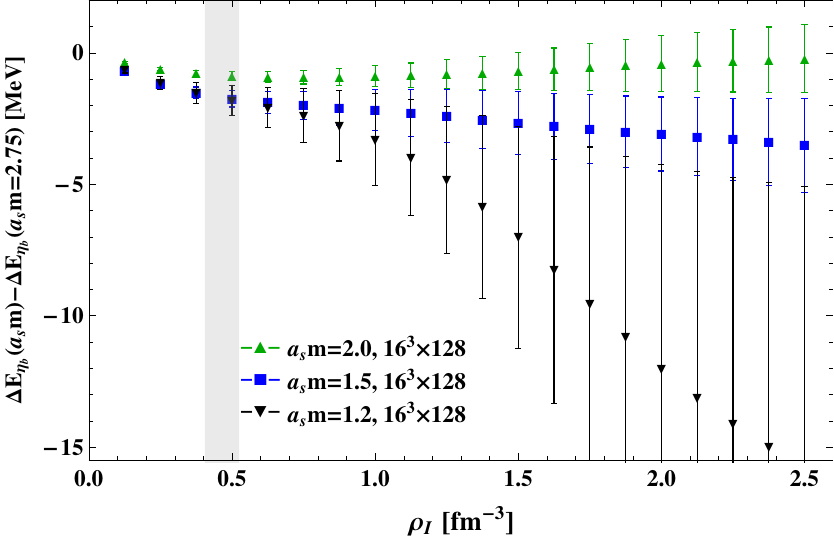}
    \\ \vspace*{5mm}
    \includegraphics[width=0.95\columnwidth]{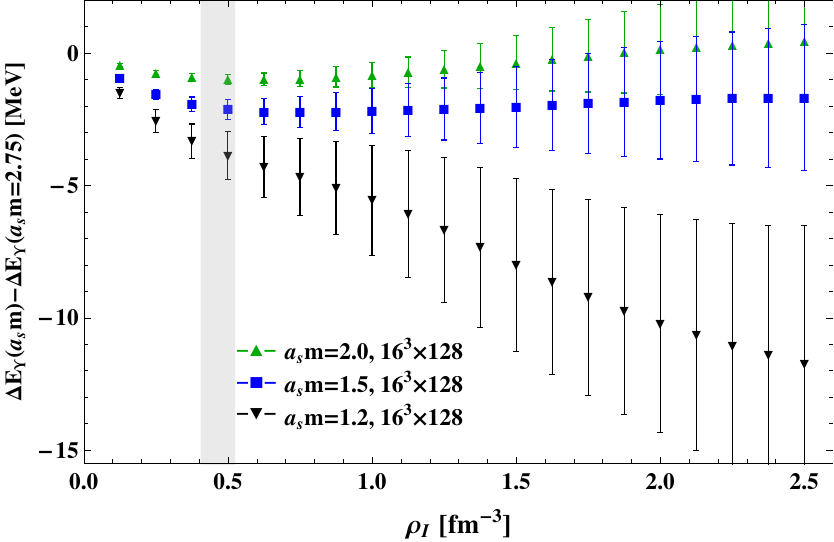}
    \caption{The difference of the $\eta_b$ (upper) and $\Upsilon$
      (lower) energy shift for a given heavy quark mass from the shift for $a_s
      m=2.75$ is shown as a function of the isospin charge density.
      Results are shown for the $16^3\times128$ ensemble.}
    \label{fig:deltaE_vs_mu_mB}
  \end{center}
\end{figure}

As discussed in Section \ref{sec:nrqcd}, we have performed
calculations for four different values of the heavy-quark mass, $a_s
m$, ranging from the bottom-quark mass down to $\sim1.5$ times the
charm-quark mass.  The analysis of the in-medium correlators and
ratios is very similar for all masses and we do not present it in
detail.  To investigate the variation of the energy shifts as a
function of the heavy-quark mass we compute
$\Delta E_{n,\bbar}(a_s
m)-\Delta E_{n,\bbar}(a_s m=2.75)$
using the bootstrap method.
  Because of correlations between the measurements
for different values of the heavy-quark mass, this provides a more
statistically precise determination of the difference than would be evident
from a naive comparison.  Figure \ref{fig:deltaE_vs_mu_mB}
shows these energy differences for the different values of $a_s m$.
It is apparent that  the strength of the energy shift in both $\eta _b$ and
$\Upsilon$ increases as the heavy-quark mass decreases, in line with expectations
from the potential model discussed above. Since the quarkonium states
for lower heavy-quark masses are physically larger, they probe regions of 
larger quark--anti-quark separation where the potential shift is more significant.

\section{Discussion}
\label{sec:discuss}

Heavy-quark bound states provide an important probe of the properties
of a medium and have been used in this work to investigate systems of large 
isospin charge density created by many-pion correlators. Specifically,
we have used lattice QCD to investigate how the presence
of this medium modifies the NRQCD
energies of various quarkonium states. Our calculations make use of
ensembles of lattices with three different physical volumes at a single
lattice spacing and at a single light quark mass corresponding to
 $m_\pi\sim390$ MeV. We have found a measurable decrease in
the energy of both the $\eta_b$ and $\Upsilon$ states and in the
spin-averaged $P$-wave energy. This decrease grows as the isospin
charge increases before flattening at a charge density at which
Ref.~\cite{Detmold:2012wc} previously observed strongly non-monotonic
behaviour of the energy density of the medium. The saturation
of the energy shift provides further support to the conjecture that a
transition from a pion gas to a Bose-Einstein condensate of pions occurs at
this point. In the region of low isospin density,
the energy shift shows an increase with the density, as expected
from a potential model augmented with the hadronic screening effect found in Ref.~\cite{Detmold:2010au},
but the effect is larger than predicted by the model.

We have investigated how the observed energy shifts depend on the
mass of the heavy quark--anti-quark pair, finding an enhanced effect
for lighter masses. Given the phenomenological interest in $J/\Psi$
suppression in medium, it will be interesting to investigate the
analogous behaviour in the charmonium sector using alternative
formulations of the heavy-quark action more appropriate for the charm
quark. However, this is beyond the scope of the current work.

A similar study of NRQCD quarkonium correlators in QC$_2$D
(two-colour QCD) at non-zero quark chemical potential was recently
presented by Hands {\it et al.} in Ref.~\cite{Hands:2012yy}.  In contrast to
QCD with three colours, in QC$_2$D, the addition of a
quark chemical potential does not result in a complex action and
numerical calculations can be performed efficiently \cite{Hands:2006ve,Hands:2007uc,Hands:2010gd}.  In
Ref.~\cite{Detmold:2012wc} it was pointed out that the phase structure
of QCD at nonzero $\mu_I$ has an intriguing similarity to that of
QC$_2$D at nonzero quark chemical potential. It is apparent that the
similarities persist to the case of quarkonium energy shifts in medium
as an at least qualitatively  similar dependence on the  charge density/chemical
potential is observed in the two-colour QCD case. Recent work 
\cite{Cherman:2011mh,Hanada:2011ju,Hanada:2012nj} has 
probed the connections between different gauge theories with non-zero 
(isospin) chemical potentials and, as the extent
of this similarity is surprising, this warrants further investigation. 

Finally, by looking at quarkonium-pion correlation functions on
three different volumes, we have extracted the $\eta_b$-$\pi$ and
$\Upsilon$-$\pi$ scattering lengths. Our results, interpolated to
the physical pion mass, are $a_{\eta_b,\pi}^{(\rm phys.)} = 0.0025(8)(6)\:\:{\rm fm}$
and $a_{\Upsilon,\pi}^{(\rm phys.)} = 0.0030(9)(7)\:\:{\rm fm}$.

\acknowledgments We thank K. Orginos and M. Savage for insightful
discussions on the topic of this work and R. Edwards and B. Jo\'{o}
for development of the {\tt qdp++} and {\tt chroma} software
suites~\cite{Edwards:2004sx}.  We acknowledge computational support
from the National Energy Research Scientific Computing Center (NERSC,
Office of Science of the US DOE, DE-AC02-05CH11231), and the NSF
through XSEDE resources provided by NICS.  This work was supported in
part by DOE grants DE-AC05-06OR23177 (JSA) and DE-FG02-94ER40818.  WD
was also supported by DOE OJI grant DE-SC0001784 and Jeffress Memorial
Trust, grant J-968.

\FloatBarrier

\end{document}